\documentclass[lettersize,journal]{IEEEtran}
\IEEEoverridecommandlockouts
\usepackage{makecell}
\usepackage{array}
\usepackage{graphicx,amssymb,amsmath}
\usepackage{multicol}
\usepackage[noadjust]{cite}
\usepackage{setspace}
\usepackage{subfigure}
\usepackage{graphicx}
\usepackage{float}
\usepackage{url}
\usepackage{stfloats}
\usepackage{amsthm,pifont}
\usepackage{flushend}
\usepackage{cases,subeqnarray}
\usepackage{bm,multirow,bigstrut}
\usepackage{amsmath, amsthm, amssymb}
\usepackage{textcomp}
\usepackage{latexsym,bm}
\usepackage{booktabs}
\usepackage{xcolor}
\usepackage{mathtools}
\usepackage{dsfont}
\usepackage{extarrows}
\usepackage{epsfig}
\usepackage{epstopdf}
\usepackage[noend]{algpseudocode}
\usepackage{algorithmicx}

\usepackage{algorithm}
\usepackage{algpseudocode}
\usepackage{amsmath}
\usepackage{pifont}
\usepackage{cite}
\usepackage{bm}
\usepackage{cleveref}
\usepackage{multicol}       
\usepackage{multirow}       
\usepackage{array}          
\usepackage{colortbl}
\usepackage{makecell}
\definecolor{crimson}{RGB}{192,0,0}         
\definecolor{navy}{RGB}{47,85,151}         
\usepackage{bbding}
\usepackage{graphicx}
\usepackage{booktabs}
\usepackage{algorithm}
\usepackage{algpseudocode}

\theoremstyle{plain}

\theoremstyle{plain}
\newtheorem{rem}{Remark}
\newtheorem{them}{Theorem}

\usepackage{amsmath}

\IEEEoverridecommandlockouts
\begin{document}

\title{Energy-Efficient SIM-assisted Communications: How Many Layers Do We Need?}

\author{Enyu Shi, Jiayi~Zhang,~\IEEEmembership{Senior Member,~IEEE}, Jiancheng An,~\IEEEmembership{Member,~IEEE}, \\ Marco Di Renzo,~\IEEEmembership{Fellow,~IEEE}, Bo Ai,~\IEEEmembership{Fellow,~IEEE}, and Chau Yuen,~\IEEEmembership{Fellow,~IEEE}
\thanks{E. Shi, J. Zhang, and B. Ai are with the State Key Laboratory of Advanced Rail Autonomous Operation, and also with the School of Electronics and Information Engineering, Beijing Jiaotong University, Beijing 100044, P. R. China. (e-mail: \{enyushi, jiayizhang, boai\}@bjtu.edu.cn).}
\thanks{M. Di Renzo is with Universit\'e Paris-Saclay, CNRS, CentraleSup\'elec, Laboratoire des Signaux et Syst\`emes, 3 Rue Joliot-Curie, 91192 Gif-sur-Yvette, France. (marco.di-renzo@universite-paris-saclay.fr), and with King's College London, Centre for Telecommunications Research -- Department of Engineering, WC2R 2LS London, United Kingdom (marco.di\_renzo@kcl.ac.uk).}
\thanks{J. An and C. Yuen are with the School of Electrical and Electronics Engineering, Nanyang Technological University, Singapore 639798 (e-mail: jiancheng\_an@163.com, chau.yuen@ntu.edu.sg).}
}
\maketitle
\begin{abstract}
The stacked intelligent metasurface (SIM), comprising multiple layers of reconfigurable transmissive metasurfaces, is becoming an increasingly viable solution for future wireless communication systems.
In this paper, we explore the integration of SIM in a multi-antenna base station for application to downlink multi-user communications, and a realistic power consumption model for SIM-assisted systems is presented.
Specifically, we focus on maximizing the energy efficiency (EE) for hybrid precoding design, i.e., the base station digital precoding and SIM wave-based beamforming. Due to the non-convexity and high complexity of the formulated problem, we employ the quadratic transformation method to reformulate the optimization problem and propose an alternating optimization (AO)-based joint precoding framework. Specifically, a successive convex approximation (SCA) algorithm is adopted for the base station precoding design. For the SIM wave-based beamforming, two algorithms are employed: the high-performance semidefinite programming (SDP) method and the low-complexity projected gradient ascent (PGA) algorithm.
In particular, the results indicate that while the optimal number of SIM layers for maximizing the EE and spectral efficiency differs, a design of 2 to 5 layers can achieve satisfactory performance for both.
Finally, numerical results are illustrated to evaluate the effectiveness of the proposed hybrid precoding framework and to showcase the performance enhancement achieved by the algorithm in comparison to benchmark schemes.
\end{abstract}


\IEEEpeerreviewmaketitle

\section{Introduction}
\IEEEPARstart{T}{he} upcoming sixth-generation (6G) network is anticipated to play a pivotal role in numerous areas of the future society, industry, and everyday life, necessitating extremely high communication standards regarding capacity, latency, reliability, and intelligence \cite{wang2023road}.
To meet these demands, next-generation network technologies such as massive multiple-input multiple-output (mMIMO) \cite{gong2023holographic}, extra-large scale/holographic MIMO \cite{10500425}, cell-free mMIMO \cite{OBETrans}, and movable antenna \cite{zhu2023modeling} have been proposed. These technologies aim to achieve time and space diversity gains by significantly increasing the number of antennas. However, this growth comes at the cost of higher energy consumption, raising concerns about energy efficiency (EE) and the sustainability of future wireless systems. As energy consumption becomes a critical issue for next-generation networks, improving the EE has become one of the key focuses of 6G communications \cite{buzzi2016survey}. In fact, the need for sustainable and energy-efficient wireless networks is crucial to support the massive expansion of connected devices and the increasing demand for data rates.

Among the candidate transceivers for green communications, reconfigurable intelligent surfaces (RISs) have been proposed as an effective solution to boost system performance while reducing energy usage and costs \cite{10556753,wu2019intelligent,di2020smart,10167480}. With a large number of low-cost nearly-passive elements, RIS is able to reflect the electromagnetic incident signals to any direction with high array gains by adjusting the phase profile of its elements \cite{ma2022cooperative}.
For example, the authors of \cite{huang2019reconfigurable} investigated the EE of RIS-assisted multiple-input single-output (MISO) systems considering the power consumption and phase quantization accuracy of the RIS elements. The results show that RIS can significantly improve the EE of the system, but blindly increasing the number of elements is not always beneficial. In \cite{you2020energy}, in addition, the authors explored the SE and EE trade-off in RIS-assisted MIMO systems and proposed an alternating optimization (AO) algorithm that can achieve a satisfactory balance between EE and SE.
Furthermore, the studies in \cite{le2021energy} and \cite{fotock2023energy} incorporated an RIS to reduce the energy consumption of CF mMIMO systems, aiming to maximize the system EE through optimized RIS phase shifts.
Albeit these significant performance improvements, RIS still has some inherent limitations. For instance, as a low-complexity, nearly-passive, and ``transparent" electronic component, RIS cannot function as a transceiver, making it incapable of achieving the spatial multiplexing capabilities of typical active multi-antenna base stations (BSs) \cite{an2021low}. Recent results have showcased the performance gains provided by RIS, in typical cellular network deployments, based on accurate system-level simulations, provided that the size of the surface if sufficiently large \cite{sihlbom2022reconfigurable}. A typical RIS node is deployed throughout the environment, and, therefore,  several solutions are under analysis to reduce the channel estimation and control overhead of RIS-assisted communications \cite{9244106}.

Besides the implementation of RIS, motivated by the rapid development of metasurface design, recent research on intelligent surfaces has explored the use of multi-layer architectures for electromagnetic (EM) wave-based signal processing \cite{doi:10.1126/science.aat8084,liu2022programmable,nerini2024physically,liu2024stacked}. These emerging solutions operate as diffractive devices and can be utilized, e.g., to implement low-complexity and low-power consumption MIMO transceivers. For example, the authors of \cite{liu2022programmable} introduced a programmable diffractive deep neural network structure that employs a multilayer metasurface array, where each meta-atom functions as a reconfigurable artificial neuron. Building on this idea, the authors of \cite{10158690} proposed a novel stacked intelligent metasurface (SIM)-enabled MIMO transceiver. This technology consists of multiple nearly-passive, programmable metasurfaces to form wireless transceivers that resemble an artificial neural network (ANN), providing significant signal processing capabilities for data encoding and decoding. By appropriately configuring a pair of SIMs, complex multi-stream precoding operations in the digital domain can be replaced by analog processing in the wave domain, which significantly reduces the computational complexity and the power consumption of a typical fully-digital transmitter.
For example, the authors of \cite{papazafeiropoulos2024performance} innovatively deployed two SIMs, one on the BS and another within the propagation environment, and designed a hybrid precoding strategy to enhance the system uplink sum rate performance. This approach provides new perspectives for the application of SIMs. Besides, the authors of \cite{nadeem2023hybrid} applied SIM for receiver combining and transmit precoding in holographic MIMO communications, showing the benefit of multi-layer metasurface devices as compared to their single-layer counterpart.
Besides, hybrid beamforming is widely used in mMIMO transceivers, which can reduce the complexity of the BS and improve performance \cite{8371237}. The authors of \cite{9467318} investigated beam selection strategies for low complexity hybrid beamforming applied to mMIMO systems. Besides, the authors of \cite{li2025stacked} investigated the design of a hybrid beamforming SIM-assisted transceiver in near-field wideband systems.
These studies have demonstrated that SIM, as a transceiver, can achieve the equivalent effect of increasing the number of transceiver antennas while enabling efficient precoding \cite{wang2024multi}. This ultimately enhances the optimization space while reducing power consumption, in agreement with the contemporary sought-after design criterion of sustainable communications.

At present, the majority of research on SIM focuses on reducing the complexity of BSs and enhancing system capacity \cite{an2024two,niu2024efficient,li2024stacked}.
For example, the authors of \cite{an2024two} have advanced the development of SIM-based transceivers by focusing on channel and direction-of-arrival estimation. Also, the authors of \cite{papazafeiropoulos2024achievable} have investigated the multiuser SIM system capacity by designing the hybrid digital and wave beamforming. Besides, the authors of \cite{li2024stacked} and \cite{shi2024harnessing} have explored the uplink spectral efficiency (SE) performance of SIM-assisted cell-free mMIMO networks by designing combining methods. The authors of \cite{shi2025downlink} have investigated the downlink performance of SIM-assisted cell-free mMIMO by jointly designing the BS power and SIM beamforming. Also, the authors of \cite{bahingayi2025scaling} have analyzed the achievable rate scaling laws of SIM-aided MIMO systems as a function of the number of metasurface layers and have proposed a hybrid optimization framework to enhance system performance. The authors of \cite{papazafeiropoulos2024WCL} have evaluated the performance of SIM by optimizing the SE in Rician fading channels by relying on statistical CSI. The results demonstrate that SIM can significantly enhance the system capacity in various communication scenarios. However, there has been limited research on the EE of SIM-assisted systems. To the best of our knowledge, only the authors of \cite{perovic2024energy} have conducted an analysis of the EE in SIM-based broadcast MIMO channels. The framework proposed in \cite{perovic2024energy} provides an important foundation for understanding energy-efficient MIMO broadcast channels, leveraging dirty paper coding and linear precoding, by utilizing a projected gradient method to optimize the SIM. However, it is known that user fairness is also a crucial consideration, as ensuring balanced performance across multiple users is essential for practical deployments. Additionally, comprehensive energy consumption models for SIM-based transceivers are necessary for a thorough analysis of the impact of various EE parameters.

Motivated by these observations, we offer a comprehensive modeling and analysis of the energy consumption of SIM-assisted multi-user MISO systems. The contributions of this work are delineated as follows:

\begin{itemize}
\item Considering the hardware architecture of an SIM-assisted BS, we develop a realistic power consumption model. Based on that, we formulate a hybrid precoding framework, i.e., joint BS antenna digital precoding and SIM wave-based beamforming design problem, to maximize the EE for all the users while considering user fairness. Specifically, we consider a hybrid digital and wave precoding architecture. Also, we account for practical constraints, including the antenna maximum transmission power and the minimum quality of service (QoS) for the users.

\item The problem is formulated as a mixed non-linear program (MNLP), where the coupling between the BS transmit power allocation and SIM beamforming results in a non-convex objective function, making it difficult to find an optimal solution. To tackle this complexity, we first apply the quadratic transform method to reformulate the optimization problem and then apply an AO framework to derive suboptimal solutions. Specifically, we utilize a successive convex approximation (SCA) algorithm to optimize the BS precoding. To optimize the SIM, we employ two distinct algorithms: the high-performance semidefinite programming (SDP) framework and the low-complexity projected gradient ascent (PGA) method. Also, we analyze the complexity of the proposed joint precoding framework.

\item We analyze the impact of various system parameters on both system EE and SE. Simulation results demonstrate that an SIM-assisted BS can significantly improve the SE and EE of communication systems. Notably, the results indicate that to ensure satisfactory performance in both EE and SE, selecting a number of SIM layers between 2-5 is an appropriate and effective choice.
Moreover, the obtained results indicate that the proposed hybrid precoding achieves a 80\% improvement in terms of EE compared to a BS digital precoding without utilizing SIM beamforming. These findings underscore the potential of SIM technology, offering a more effective approach for optimizing the energy and spectral efficiency of future wireless communication systems.
\end{itemize}

The remainder of this paper is structured as follows.
Section \uppercase\expandafter{\romannumeral2} outlines the SIM-assisted mult-user MISO system model, including the channel model, downlink data transmission, power consumption model, and problem formulation.
Next, Section \uppercase\expandafter{\romannumeral3} proposes the joint precoding framework, consisting of the BS antenna precoding and SIM wave-based beamforming.
Then, numerical results are illustrated and discussed in Section \uppercase\expandafter{\romannumeral4}.
Finally, Section \uppercase\expandafter{\romannumeral5} concludes this paper.

\textbf{Notation:} Column vectors and matrices are denoted by boldface lowercase letters $\mathbf{x}$ and boldface uppercase letters $\mathbf{X}$, respectively. The superscripts $\mathbf{x}^{\rm{H}}$, $x^\mathrm{T}$, and $x^\mathrm{*}$ represent the conjugate, transpose, and conjugate transpose, respectively. The notation $\triangleq$, $\left\|  \cdot  \right\|$, and $\left\lfloor  \cdot  \right\rfloor $ indicate the definitions, the Euclidean norm, and the floor function, respectively. ${\rm{tr}}\left(  \cdot  \right)$, $\mathbb{E}\left\{  \cdot  \right\}$, and ${\rm{Cov}}\left\{  \cdot  \right\}$ denote the trace, expectation, and covariance operators, respectively. ${\text{diag}}\left( {{a_1}, \cdots ,{a_n}} \right)$ denotes a diagonal matrix.
A circularly symmetric complex Gaussian random variable $x$ with mean $0$ and variance $\sigma^2$ is denoted by $x \sim \mathcal{C}\mathcal{N}\left( {0,{\sigma^2}} \right)$. $\nabla$ denotes the gradient operator. $\mathbb{B}^n$, $\mathbb{Z}^n$, $\mathbb{R}^n$, and $\mathbb{C}^n$ represent the $n$-dimensional spaces of binary, integer, real, and complex numbers, respectively. Finally, the $N \times N$ zero matrix and identity matrix are denoted by $\mathbf{0}_{N}$ and $\mathbf{I}_{N}$, respectively.


\section{System Model}\label{se:model}
\begin{figure}[t]
\centering
\includegraphics[scale=0.65]{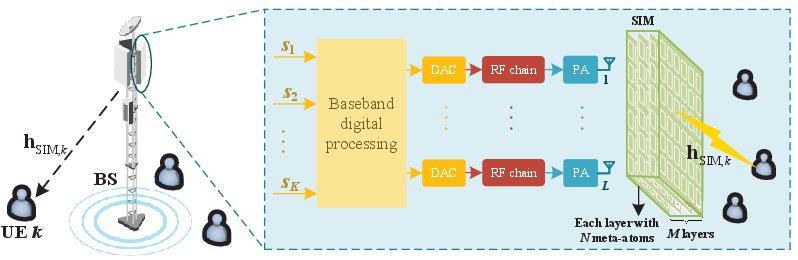}\vspace{-0.2cm}
\caption{Illustration of the considered SIM-based multi-user MISO system with hybrid digital and wave-domain precoding.} \label{system_model}\vspace{-0.2cm}
\end{figure}

As shown in Fig.~\ref{system_model}, we consider an SIM-based downlink multi-user MISO system with hybrid digital and wave precoding architecture, which consists of an SIM-assisted BS and $K$ UEs. We assume that the BS and each UE are equipped with $L$ antennas and a single antenna, respectively. We assume that the number of antennas $L$ is greater than or equal to the number of UEs $K$, i.e., $L \ge K$. Also, we assume that the SIM has $M$ reconfigurable metasurface layers and $N$ meta-atoms in each layer. Let ${\cal K} = \{ 1, \ldots ,K\}$, ${\cal L} = \{ 1, \ldots ,L\}$, ${\cal M} = \{ 1, \ldots ,M\}$, and ${\cal N} = \{ 1, \ldots ,N\}$ denote the index sets of UEs, BS antennas, SIM metasurface layers, and meta-atoms per layer, respectively. Furthermore, the SIM is connected to an intelligent controller at the BS, capable of applying a distinct and tunable phase shift to the EM waves passing through each meta-atom \cite{an2023stacked2}. By properly adjusting the phase shifts in each metasurface and the digital precoding of the BS, proposed SIM-assisted BS implements the downlink beamforming in the hybrid digital and EM wave domain \cite{papazafeiropoulos2024achievable}.

\subsection{Channel Model}
We use ${e^{j\varphi _{m}^n}},\forall m \in {\cal M},\forall n \in {\cal N}$ with  $\varphi _{m}^n \in \left[ {0,2\pi } \right)$ denoting the $n$-th meta-atom's phase shift of the $m$-th metasurface layer at the SIM. Hence, the diagonal phase shift matrix ${{\bf{\Phi }}_{m}}$ for the $m$-th metasurface layer at the SIM can be denoted as ${{\bf{\Phi }}_{m}} = {\rm{diag}}( {{e^{j\varphi _{m}^1}},{e^{j\varphi _{m}^2}}, \ldots ,{e^{j\varphi _{m}^N}}} ) \in \mathbb{C} {^{N \times N}},\forall m \in {\cal M}$. Furthermore, let ${\bf{W}}_{1} = {[ {{\bf{w}}_{1,1}, \ldots ,{\bf{w}}_{1,L}^{}} ]} \in \mathbb{C} {^{N \times L}}$ denote the transmission matrix from the BS antennas to the first metasurface layer of the SIM, where ${{\bf{w}}_{1,l}^{{}}} \in \mathbb{C} {^{N}}$ denotes the transmission vector from the $l$-th antenna of the BS to the first metasurface layer of the SIM. Let ${{\bf{W}}_{m}} \in \mathbb{C} {^{N \times N}},\forall m \ne 1,\forall m \in {\cal M}$ denote the transmission matrix from the $(m-1)$-th to the $m$-th metasurface layer of SIM. According to the Rayleigh-Sommerfeld diffraction theory \cite{lin2024stacked}, the $\left( {n,n'} \right)$-th element is expressed as

\begin{align}\label{w}
w_{m}^{n,n'} = \frac{{{d_x}{d_y}\cos \chi _{m}^{n,n'}}}{{d_{m}^{n,n'}}}\left( {\frac{1}{{2\pi d_{m}^{n,n'}}} - j\frac{1}{\lambda }} \right){e^{j2\pi \frac{{d_{m}^{n,n'}}}{\lambda }}},
\end{align}
where $\lambda$ represents the carrier wavelength, $d_{m}^{n,n'}$ indicates the corresponding transmission distance which is determined by \cite[Eq.~(26)]{an2023stacked2}. $d_{x} \times d_{y}$ indicates the size of each SIM meta-atom, and ${\chi _{m}^{n,n'}}$ represents the angle between the propagation direction and the normal direction of the $(m-1)$-th metasurface layer of the SIM. Similarly, the $n$-th element $w_{1,l}^n$ of ${{\bf{w}}_{1,l}^{{}}}$ can be obtained from \eqref{w}.
Hence, the wave-based beamforming matrix ${\mathbf{G}} \in \mathbb{C}{^{N \times N}}$ of the BS, enabled by SIM, is obtained as
\begin{align}\label{G_l}
{{\bf{G}}} = {{\bf{\Phi }}_{M}}{{\bf{W}}_{M}}{{\bf{\Phi }}_{M - 1}}{{\bf{W}}_{M - 1}} \ldots {{\bf{\Phi }}_{2}}{{\bf{W}}_{2}}{{\bf{\Phi }}_{1}}.
\end{align}

Then, we consider a quasi-static flat-fading channel model.
Let ${\bf{h}}_{{\rm{SI}}{{\rm{M}}},k} \in \mathbb{C}{^{N \times 1}}$ denote the direct channel between the last metasurface layer of the SIM to UE $k$. Specifically, we assume that ${\bf{h}}_{{\rm{SI}}{{\rm{M}}},k}$ is a spatially correlated Rayleigh fading channel vector with  ${\bf{h}}_{{\rm{SI}}{{\rm{M}}},k} \sim {\cal C}{\cal N}\left( {0,{{\bf{R}}_{{\rm{SI}}{{\rm{M}}},k}}} \right)$.\footnote{We consider a scenario with abundant scatterers, typical of dense urban environments, where the direct line-of-sight (LoS) component is weak such that Rayleigh fading is a more suitable channel model compared to Rician fading. The proposed optimization frameworks are, however, independent of the specific statistical channel model being considered.} ${{\bf{R}}_{{\rm{SI}}{{\rm{M}}},k}} = {\beta _{{\rm{SI}}{{\rm{M}}},k}}{\bf{R}} \in \mathbb{C}{^{N \times N}}$ where ${\beta _{{\rm{SI}}{{\rm{M}}},k}}$ denotes the distance-dependent path loss between UE $k$ and the SIM, and the covariance matrix ${\bf{R}} \in \mathbb{C}{^ {N \times N}}$ describes the spatial correlation among different meta-atoms of the last metasurface layer of the SIM \cite{gong2024holographic}. Considering an isotropic scattering environment with multipath components uniformly distributed, the $\left( {n,n'} \right)$-th element of $\mathbf{R}$ is ${{\bf{R}}_{n,n'}} = {\rm{sinc}}\left( {2{{{d_{n,n'}}} \mathord{\left/
 {\vphantom {{{d_{n,n'}}} \lambda }} \right.
 \kern-\nulldelimiterspace} \lambda }} \right)$ \cite{bjornson2020rayleigh}, where $d_{n,n'}$ denotes the inter-distance between two adjacent meta-atoms on the same metasurface layer, and ${\rm{sinc}}\left( x \right) = {{\sin \left( {\pi x} \right)} \mathord{\left/
 {\vphantom {{\sin \left( {\pi x} \right)} {\left( {\pi x} \right)}}} \right.
 \kern-\nulldelimiterspace} {\left( {\pi x} \right)}}$ denotes the normalized sinc function. Then, the total channel ${{\bf{h}}_{k}} \in \mathbb{C}{^{L \times 1}}$ from UE $k$ to the BS can be denoted as
\begin{align}\label{h_k}
{{\bf{h}}_{k}} &= {\bf{W}}_{1}^{\rm{H}}{\bf{G}}^{\rm{H}}{\bf{h}}_{{\rm{SI}}{{\rm{M}}},k}.
\end{align}
\begin{rem}
In \eqref{h_k}, we observe that the SIM modifies the channel state by adjusting the transmission coefficients of the SIM metasurface layers. Unlike the uncontrollable direct channels in traditional MIMO systems, this channel is affected by signals aggregated from multiple metasurface layers of the SIM, which provides the opportunity to orthogonalize the channels associated with multiple users as the EM waves propagate through the SIM. This enlarges the optimization space of SIM-aided MIMO systems as compared to traditional MIMO systems.
\end{rem}

\subsection{Downlink Data Transmission}
In the downlink, the BS adopts digital precoding for signal processing. Then, the downlink transmitted signal ${{\bf{x}}} \in \mathbb{C}{^{L \times 1}}$ at the BS can be written as
\begin{align}\label{x}
    {\bf{x}} = \sum\limits_{k = 1}^K {{{\bf{p}}_k}{s_k}},
\end{align}
where ${{\bf{p}}_k} = \left[ {{p_1}, \ldots ,{p_L}} \right] \in \mathbb{C}{^{L \times 1}}$ represents the transmission digital precoding vector of UE $k$ and $s_k$ is the information symbol intended to UE $k$.
Then, the received signal ${y_k} \in \mathbb{C}$ at UE $k$ can be expressed as \cite{chen2023channel}
\begin{align}\label{y_k}
&{y_k} = {\bf{h}}_{{\rm{SIM}},k}^{\rm{H}}{\bf{G}}{{\bf{W}}_1}{\bf{x}} + {n_k} = {\bf{h}}_{{\rm{SIM}},k}^{\rm{H}}{\bf{G}}{{\bf{W}}_1}\sum\limits_{k = 1}^K {{{\bf{p}}_k}{s_k}}  + {n_k}\notag\\
&= {\bf{h}}_{{\rm{SIM}},k}^{\rm{H}}{\bf{G}}{{\bf{W}}_1}{{\bf{p}}_k}{s_k} + \mathop \sum \limits_{j = 1,j \ne k}^K {\bf{h}}_{{\rm{SIM}},k}^{\rm{H}}{\bf{G}}{{\bf{W}}_1}{{\bf{p}}_j}{s_j} + {n_k},
\end{align}
where ${n_k} \sim {\cal CN}(0,\sigma_k^2)$ denotes the additive noise at UE $k$. Note that the first term on the right-hand side of \eqref{y_k} is the desired signal to UE $k$, while the second term denotes the interference from other UEs.

\begin{rem}
Differently from previous studies that did not consider digital precoding to reduce hardware complexity \cite{10158690}, we adopt a hybrid precoding architecture that integrates digital precoding at the BS with SIM-enhanced wave-based beamforming. This approach offers effective multi-user interference management and provides greater beamforming flexibility, allowing the BS to serve more users while optimizing both SE and EE.
\end{rem}

\subsection{Total Power Consumption Model}
The total power consumed by the considered SIM-based system consists of the traditional BS (excluding SIM) power, the hardware static power consumed by the mobile users, and the SIM power consumption.
Then, the total power consumption of the considered SIM-based system can be expressed by
\begin{align}\label{P_total}
{P_{{\rm{total}}}} = {P_{{\rm{BS}}}} + {P_{{\rm{SIM}}}} + K{P_{{\rm{UE}}}},
\end{align}
where $P_{\rm{UE}}$ is the fixed power consumption of each UE device,\footnote{In this study, we focus on BS-side EE optimization and model the UE power consumption using a constant power term, as considered in prior works \cite{huang2019reconfigurable} and \cite{zhang2023energy}. A more general EE evaluation accounting for the power consumption at the UE side is an important direction for future research.} and $P_{\rm{BS}}$ and $P_{\rm{SIM}}$ are the power consumption of the BS and SIM, which are computed as follows.
\subsubsection{Power Consumption of BS}
The total power consumption ${P_{{\rm{BS}}}}$ at the BS depends on the state (i.e., active or inactive) of each antenna, which comprises two parts, i.e., the hardware power consumption of the RF chains (denoted as $P_{\rm{RF}}$) and the static power consumption (denoted as $P_{\rm{S}}$). Thus, the total power consumption at the BS is given by \cite{tervo2018energy}
\begin{align}\label{P_BS}
{P_{{\rm{BS}}}} = {P_{\rm{S}}} + {P_{{\rm{RF}}}},
\end{align}
where ${P_{\rm{S}}}$ is a constant, and ${P_{{\rm{RF}}}} = {P_{{\rm{Act}}}} + {P_{{\rm{PA}}}}$ is computed as follows.
Let $P^{\rm{active}}_{\rm{RF}}$ denote the static power consumption of an active RF chain, while no power is required for an inactive RF chain. Thus, the power consumption due to all selected active RF chains is expressed as \cite{zhang2023energy}
\begin{align}\label{P_RF}
{P_{{\rm{Act}}}} = P_{{\rm{RF}}}^{{\rm{active}}}\sum\limits_{l = 1}^L {\rm{I}} \left( {\sum\limits_{k = 1}^K {{{\left| {{\bf{e}}_l^{\rm{T}}{{\bf{p}}_k}} \right|}^2}} } \right),
\end{align}
where ${\rm{I}}\left(  \cdot  \right)$ represents the indicator function. We use ${\rm{I}}\left(  \cdot  \right)$ to indicate whether the RF chain is activated. If the RF chain is activated, we have ${\rm{I}}\left(  \cdot  \right) = 1$; if it is not activated we have ${\rm{I}}\left(  \cdot  \right) = 0$.
In this work, we consider a fully digital precoding architecture at the BS, which implies that the number of RF chains is equal to the number of antennas. Under this setup, the RF chain selection is equivalent to BS antenna selection.

Let $\eta_l$ and $P^{out}_l$ denote the PA efficiency and the transmit power of antenna $l$, respectively. Then, the power consumption of the PA connected to antenna $l$ is given by
\begin{align}\label{P_PA1}
P_{{\rm{PA}}}^l = {{P_l^{out}} \mathord{\left/
 {\vphantom {{P_l^{out}} {{\eta _l}}}} \right.
 \kern-\nulldelimiterspace} {{\eta _l}}} = \sum\limits_{k = 1}^K {{{{{\left| {{\bf{e}}_l^{\rm{T}}{{\bf{p}}_k}} \right|}^2}} \mathord{\left/
 {\vphantom {{{{\left| {{\bf{e}}_l^{\rm{T}}{{\bf{p}}_k}} \right|}^2}} {{\eta _l}}}} \right.
 \kern-\nulldelimiterspace} {{\eta _l}}}},
\end{align}
where ${{\bf{e}}_l} \in \mathbb{R}{^l}$ denotes the $l$-column of the identity matrix $\mathbf{I}_l \in \mathbb{R}{^{L\times L}}$. Without loss of generality, the PA efficiencies are identically set as $\eta_l = \eta, l \in \cal{L}$. Therefore, the power consumption of the PAs is finally expressed as\footnote{The static power consumption of the PAs is incorporated into the considered power consumption model, $P_{\rm{S}}$, along with other static power components \cite{tervo2018energy}.}

\begin{align}\label{P_PA2}
{P_{{\rm{PA}}}} = \sum\limits_{l = 1}^L {P_{{\rm{PA}}}^l}  = \sum\limits_{k = 1}^K {{{{{\left\| {{{\bf{p}}_k}} \right\|}^2}} \mathord{\left/
 {\vphantom {{{{\left\| {{{\bf{p}}_k}} \right\|}^2}} \eta }} \right.
 \kern-\nulldelimiterspace} \eta }} .
\end{align}

\begin{rem}
    According to \eqref{P_RF} and \eqref{P_PA1}, the RF chain $l$ is activated only when the digital precoding power of antenna $l$, i.e., $P^{out}_l$, reaches a certain threshold ${\rho _l}$. Using the approximation for the indicator function detailed next in \eqref{I(x)}, we assume that the RF chain $l$ is activated if $x \ge 10^{-4}$ in ${\text{I}}\left( x \right)$.\footnote{${\text{I}}\left( x \right)$ is an approximate step function that jumps to a value close to 1 when the variable reaches a very small value, such as $x \ge 10^{-4}$.} In this work, as mentioned, we consider a fully digital precoding at the BS, where the number of RF chains is equal to the number of BS antennas.
    The authors of \cite{perovic2024energy} provided valuable insights into the impact of reducing the number of RF chains for EE optimization, demonstrating that minimizing the number of active RF chains can effectively enhance the EE of SIM-assisted communication systems.
\end{rem}

\newcounter{mytempeqncnt}
\begin{figure*}[t!]
\normalsize
\setcounter{mytempeqncnt}{1}
\setcounter{equation}{14}
\begin{align}\label{EE}
{\rm{EE = }}\frac{{{\rm{BW}} \times \sum\limits_{k = 1}^K {{{\log }_2}\left( {1 + {\gamma _k}} \right)} }}{{{P_{\rm{S}}} + \sum\limits_{k = 1}^K {{{{{\left\| {{{\bf{p}}_k}} \right\|}^2}} \mathord{\left/
 {\vphantom {{{{\left\| {{{\bf{p}}_k}} \right\|}^2}} \eta }} \right.
 \kern-\nulldelimiterspace} \eta }}  + P_{{\rm{RF}}}^{{\rm{active}}}\sum\limits_{l = 1}^L {\rm{I}} \left( {\sum\limits_{k = 1}^K {{{\left| {{\bf{e}}_l^{\rm{T}}{{\bf{p}}_k}} \right|}^2}} } \right) + NM{P_{{\rm{meta}}}} + {P_{{\rm{controller}}}} + K{P_{{\rm{UE}}}}}}.
\end{align}
\setcounter{equation}{15}
\hrulefill
\end{figure*}

\subsubsection{Power Consumption of SIM}
The power consumption of the SIM is divided into the fixed power of the controller ${P_{{\rm{controller}}}}$ and the power of each meta-atom $P_{\rm{meta}}$. Then, the total power consumption of the SIM ${P_{{\rm{SIM}}}}$ can be expressed by
\begin{align}\label{P_SIM}
\setcounter{equation}{10}
{P_{{\rm{SIM}}}} = NM{P_{{\rm{meta}}}} + {P_{{\rm{controller}}}}.
\end{align}
Finally, we can obtain the total power consumption of the considered SIM-based MISO system as
\begin{align}\label{P_total_final}
{P_{{\rm{total}}}} &= {P_{\rm{S}}} + \sum\limits_{k = 1}^K {{{{{\left\| {{{\bf{p}}_k}} \right\|}^2}} \mathord{\left/
 {\vphantom {{{{\left\| {{{\bf{p}}_k}} \right\|}^2}} \eta }} \right.
 \kern-\nulldelimiterspace} \eta }}  + P_{{\rm{RF}}}^{{\rm{active}}}\sum\limits_{l = 1}^L {\rm{I}} \left( {\sum\limits_{k = 1}^K {{{\left| {{\bf{e}}_l^{\rm{T}}{{\bf{p}}_k}} \right|}^2}} } \right)\notag\\
 &+ NM{P_{{\rm{meta}}}} + {P_{{\rm{controller}}}} + K{P_{{\rm{UE}}}}.
\end{align}

\subsection{Problem Fomulation}
Based on the considered system model, we consider maximizing the EE of the proposed SIM-based multi-user MISO system by jointly optimizing the BS digital precoding and SIM wave-based beamforming matrix. To this end, the signal-to-interference-plus-noise ratio (SINR) for the transmitted symbol $s_k$ at UE $k$ can be calculated by using \eqref{y_k} as
\begin{align}\label{gamma_k}
{\gamma _k} = \frac{{{{\left| {{\bf{h}}_{{\rm{SIM}},k}^{\rm{H}}{\bf{G}}{{\bf{W}}_1}{{\bf{p}}_k}} \right|}^2}}}{{\mathop \sum \limits_{j = 1,j \ne k}^K {{\left| {{\bf{h}}_{{\rm{SIM}},k}^{\rm{H}}{\bf{G}}{{\bf{W}}_1}{{\bf{p}}_j}} \right|}^2} + \sigma _k^2}}.
\end{align}

Thereby, the sum rate of all UEs is given by
\begin{align}\label{R_sum}
{R_{\rm{sum}}} = {\sum\limits_{k = 1}^K} R_k = \sum\limits_{k = 1}^K {{{\log }_2}\left( {1 + {\gamma _k}} \right)},
\end{align}
where $R_k$ is the rate of UE $k$. Then, the EE of the considered system is defined as ${\rm{EE  \buildrel \Delta \over =  }}\frac{{{\rm{BW}} \times {R_{\rm{sum}}}}}{{{P_{{\rm{total}}}}}}$, with BW denoting the transmission bandwidth, which is expressed in \eqref{EE} at the top of this page.

Finally, the EE maximization problem can be formulated as
\setcounter{equation}{15}
\begin{subequations}\label{P_0}
  \begin{align}
\mathcal{P}^0: &\mathop {{\rm{max}}}\limits_{{{\bf{p}}_k},\left\{ {{{\bf{\Phi }}_m}} \right\}} \quad {\rm{EE = }}\frac{{{\rm{BW}} \times \sum\limits_{k = 1}^K {{R_k}\left( {\left\{{{\bf{p}}_k}\right\},\left\{ {{{\bf{\Phi }}_m}} \right\}} \right)} }}{{{P_{{\rm{total}}}}}}\label{0-1}\\
&{\rm{s}}{\rm{.t}}{\rm{.}}\quad \quad \;\;\;{R_k}\left( {\left\{{{\bf{p}}_k}\right\},\left\{ {{{\bf{\Phi }}_m}} \right\}} \right) \ge R_k^{\min },\quad \forall k \in {\cal K},\label{0-2}\\
&\quad \quad \quad \;\;\;\sum\limits_{k = 1}^K {{{\left\| {{{\bf{p}}_k}} \right\|}^2}}  \le {P_{{\rm{max}}}},\label{0-3}\\
&\quad \quad \quad \;\;\;\sum\limits_{k = 1}^K {{{\left| {{\bf{e}}_l^{\rm{T}}{{\bf{p}}_k}} \right|}^2}}  \le P_l^{\max },\quad \forall l \in {\cal L},\label{0-4}\\
&\quad \quad \quad \;\;\;\varphi _m^n \in \left[ {0,2\pi } \right),\quad \forall n \in {\cal N},\forall m \in {\cal M},\label{0-5}
  \end{align}
\end{subequations}
where $R_k^{\min }$ denotes the required data rate of UE $k$, ${P_{\max }}$ denotes the maximum transmission power at the BS, and $P_l^{\max}$ denotes the maximum transmit power of antenna $l$ which is less than the maximum transmission power ${P_{\max}}$ by definition. The constraint \eqref{0-2} ensures that the rate of each UE is not less than the required minimum rate. Then, \eqref{0-3} and \eqref{0-4} guarantee the transmit power constraints of the BS and each antenna. Finally, \eqref{0-5} accounts for the phase shifts applied by the meta-atoms, resulting in a unit-modulus constraint.

Note that the complex objective function \eqref{0-1} is non-convex, and the constant modulus constraint on the SIM transmission coefficients in \eqref{0-4} is also non-convex. Additionally, the SIM phase shifts and BS precoding coefficients are highly intertwined in the objective function, making it non-trivial to obtain the optimal solution for problem \eqref{P_0}. As such, in Section \ref{AO}, we provide an efficient algorithm for solving the formulated non-convex problem.

\section{Proposed Hybrid Precoding Framework} \label{AO}
In this section, we present the proposed hybrid precoding framework to solve the EE optimization problem ${\cal{P}}^0$.
Specifically, the section is structured as follows. An overview of the proposed framework is first provided in Subsection \ref{Overview}, where the problem ${\cal{P}}^0$ is divided into three subproblems. Then, the algorithms to solve the subproblems are given in Subsection \ref{opt_t}, \ref{opt_p}, and \ref{opt_phase}, respectively.

\subsection{Overview of the Proposed Optimization Framework}\label{Overview}
As the foundation for the hybrid precoding design, we assume that the BS acquires perfect CSI of the entire SIM-based multi-user MISO system in advance.
The objective function in this problem is non-convex due to its fractional structure. To address this, we apply the quadratic transformation method by introducing an auxiliary variable $t \in \mathbb{R}$, resulting in the following objective function \cite{shen2018fractional}
\begin{align}\label{f}
f\left( {\left\{{{\bf{p}}_k}\right\},\left\{ {{{\bf{\Phi }}_m}} \right\},t} \right) &= 2t{\left( {\sum\limits_{k = 1}^K {{R_k}\left( {\left\{{{\bf{p}}_k}\right\},\left\{ {{{\bf{\Phi }}_m}} \right\}} \right)} } \right)^{\frac{1}{2}}} \notag\\
&- {t^2}{P_{{\rm{total}}}}(\left\{{\mathbf{p}_k}\right\})    ,
\end{align}
It is not difficult to show that $f\left( {\left\{{{\bf{p}}_k}\right\},\left\{ {{{\bf{\Phi }}_m}} \right\},t} \right) \le {{\sum\limits_{k \in {\cal K}} {{R_k}\left( {\left\{{{\bf{p}}_k}\right\},\left\{ {{{\bf{\Phi }}_m}} \right\}} \right)} } \mathord{/
 {\vphantom {{\sum\limits_{k \in {\cal K}} {{R_k}\left( {{{\bf{p}}_k},\left\{ {{{\bf{\Phi }}_m}} \right\}} \right)} } {{P_{{\rm{total}}}}}}}
 \kern-\nulldelimiterspace} {{P_{{\rm{total}}}}}}$ and the equality holds when $t = {{{{\left( {\sum\limits_{k = 1}^K {{R_k}\left( {\left\{{{\bf{p}}_k}\right\},\left\{ {{{\bf{\Phi }}_m}} \right\}} \right)} } \right)}^{\frac{1}{2}}}} \mathord{/
 {\vphantom {{{{\left( {\sum\limits_{k \in {\cal K}} {{R_k}\left( {{{\bf{p}}_k},\left\{ {{{\bf{\Phi }}_m}} \right\}} \right)} } \right)}^{\frac{1}{2}}}} {{P_{{\rm{total}}}}}}}
 \kern-\nulldelimiterspace} {{P_{{\rm{total}}}}}}$. Therefore, the problem ${\mathcal{P}}^0$ is equivalent to ${\mathcal{P}}^1$ as
\begin{subequations}\label{P_1}
  \begin{align}
{\mathcal{P}}^1:
&\mathop {{\rm{max}}}\limits_{{{\bf{p}}_k},\left\{ {{{\bf{\Phi }}_m}} \right\},t} \quad f\left( {\left\{{{\bf{p}}_k}\right\},\left\{ {{{\bf{\Phi }}_m}} \right\},t} \right)\label{1-1}\\
&\quad\;\;{\rm{s}}{\rm{.t}}{\rm{.}}\quad \;\;\;\;{\gamma _k} \ge \gamma _k^{\min },\quad \forall k \in {\cal K},\label{1-2}\\
&\quad\;\;\;\quad \quad \quad \eqref{0-3},\eqref{0-4},\eqref{0-5},\notag
\end{align}
\end{subequations}
where $\gamma _k^{\min } = {2^{R_k^{\min }}} - 1$.

However, the problem in \eqref{P_1} is still not convex due to the coupling among the variables and the non-convex constraints. To solve problem ${\cal{P}}^1$, we propose an AO-based joint BS digital precoding and SIM wave-based beamforming framework by iteratively optimizing the auxiliary variables $t$, the digital transmit precoding ${{{\bf{p}}_k}}$, and the SIM beamforming matrices ${{{\bf{\Phi }}_m}}$. 

\subsection{Optimization of $t$ Given $\left\{ {{{\bf{p}}_k},{{\bf{\Phi }}_m}} \right\}$ }\label{opt_t}
It is observed that the objective function $f\left( {\left\{{{\bf{p}}_k}\right\},\left\{ {{{\bf{\Phi }}_m}} \right\},t} \right)$ is concave with respect to (w.r.t) $t$ given $\left\{ {{{\bf{p}}_k}, {{{\bf{\Phi }}_m}} } \right\}$. Therefore, a closed-form solution $t^*$ can be obtained through the application of the partial derivative approach, i.e., ${{{\partial f\left( {\left\{{{\bf{p}}_k}\right\},\left\{ {{{\bf{\Phi }}_m}} \right\},t} \right)} \mathord{\left/
 {\vphantom {{\partial f\left( {{{\bf{p}}_k},\left\{ {{{\bf{\Phi }}_m}} \right\},t} \right)} {\left. {\partial t} \right|}}} \right.
 \kern-\nulldelimiterspace} {\left. {\partial t} \right|}}_{t = {t^ * }}} = 0$, and $t^*$ can be obtained by
 \begin{align}\label{t}
 t^* = {{{{\left( {\sum\limits_{k = 1}^K {{R_k}\left( {\left\{{{\bf{p}}_k}\right\},\left\{ {{{\bf{\Phi }}_m}} \right\}} \right)} } \right)}^{\frac{1}{2}}}} \mathord{/
 {\vphantom {{{{\left( {\sum\limits_{k \in {\cal K}} {{R_k}\left( {{{\bf{p}}_k},\left\{ {{{\bf{\Phi }}_m}} \right\}} \right)} } \right)}^{\frac{1}{2}}}} {{P_{{\rm{total}}}}}}}
 \kern-\nulldelimiterspace} {{P_{{\rm{total}}}}}}
 \end{align}
By substituting $t^*$ into $f$ in \eqref{f}, one can notice that the objective function is still jointly influenced by the variables ${\bf{p}}_k$ and ${{{\bf{\Phi }}_m}}$. 

\subsection{Optimization of $\left\{{\bf{p}}_k\right\}$ Given $\left\{ {{{\bf{\Phi }}_m},t} \right\}$ }\label{opt_p}
For ease of representation, we define ${\bf{h}}_k^{\rm{H}} \buildrel \Delta \over = {\bf{h}}_{{\rm{SIM}},k}^{\rm{H}}{\bf{G}}{{\bf{W}}_1}$. Given $\left\{ {{{\bf{\Phi }}_m},t} \right\}$, problem ${\cal{P}}^1$ can be rewritten as
\begin{subequations}\label{P_BS}
  \begin{align}
{\cal{P}}^{\rm{BS}}:&\mathop {{\rm{max}}}\limits_{\left\{ {{{\bf{p}}_k}} \right\}} \quad 2t{\left( {\sum\limits_{k = 1}^K {{{\log }_2}\left( {1 + \frac{{{{\left| {{\bf{h}}_k^{\rm{H}}{{\bf{p}}_k}} \right|}^2}}}{{\mathop \sum \limits_{j = 1,j \ne k}^K {{\left| {{\bf{h}}_k^{\rm{H}}{{\bf{p}}_j}} \right|}^2} + \sigma _k^2}}} \right)} } \right)^{\frac{1}{2}}} \notag\\
 &- {t^2}\left( {\sum\limits_{k = 1}^K {{{{{\left\| {{{\bf{p}}_k}} \right\|}^2}} \mathord{\left/
 {\vphantom {{{{\left\| {{{\bf{p}}_k}} \right\|}^2}} \eta }} \right.
 \kern-\nulldelimiterspace} \eta }}  + P_{{\rm{RF}}}^{{\rm{active}}}\sum\limits_{l = 1}^L {\rm{I}} \left( {\sum\limits_{k = 1}^K {{{\left| {{\bf{e}}_l^{\rm{T}}{{\bf{p}}_k}} \right|}^2}} } \right)} \right.\notag\\
&\left. { + {P_{\rm{S}}} + NM{P_{{\rm{meta}}}} + {P_{{\rm{controller}}}} + K{P_{{\rm{UE}}}}} \right)\label{2-1}\\
&{\rm{s}}{\rm{.t}}{\rm{.}}\quad \;\;\;\frac{{{{\left| {{\bf{h}}_k^{\rm{H}}{{\bf{p}}_k}} \right|}^2}}}{{\mathop \sum \limits_{j = 1,j \ne k}^K {{\left| {{\bf{h}}_k^{\rm{H}}{{\bf{p}}_j}} \right|}^2} \!\!+\! \sigma _k^2}} \!\ge\! \gamma _k^{\min },\;\forall k \in {\cal K},\label{2-2}\\
&\quad \quad \quad \sum\limits_{k = 1}^K {{{\left\| {{{\bf{p}}_k}} \right\|}^2}}  \le {P_{{\rm{max}}}},\label{2-3}\\
&\quad \quad \quad \sum\limits_{k = 1}^K {{{\left| {{\bf{e}}_l^{\rm{T}}{{\bf{p}}_k}} \right|}^2}}  \le P_l^{\max },\quad \forall l \in {\cal L}. \label{2-4}
\end{align}
\end{subequations}
The objective function in \eqref{2-1} is non-concave because of the presence of the indicator function ${\rm{I}}\left(  \cdot  \right)$ and the fractional components, specifically the SINR terms. To address the fractional terms, auxiliary variables $\left\{ {{\gamma _k}} \right\}$ are introduced, replacing the first term in the objective function with $2t{\left( {\sum\limits_{k = 1}^K {{{\log }_2}\left( {1 + {\gamma _k}} \right)} } \right)^{\frac{1}{2}}}$. Additionally, the constraints on $\left\{ {{{\bf{p}}_k},{\gamma _k}} \right\}$ are formulated based on the UEs QoS requirements in \eqref{2-2} as
\begin{align}\label{p_gamma}
&{\left| {{\bf{h}}_k^{\rm{H}}{{\bf{p}}_k}} \right|^2} \ge {\gamma _k}\left( {\mathop \sum \limits_{j = 1,j \ne k}^K {{\left| {{\bf{h}}_k^{\rm{H}}{{\bf{p}}_j}} \right|}^2} + \sigma _k^2} \right),\quad \forall k \in {\cal K},\\
&{\gamma _k} \ge \gamma _k^{\min },\quad \forall k \in {\cal K}.
\end{align}
Based on the approach outlined in \cite{shi2011iteratively}, we select the solution set $\{ {{{\bf{p}}_k}} \}$ to meet the following additional constraints without compromising optimality as
\begin{align}\label{Re-Im}
    {\mathop{\rm Re}\nolimits} \left\{ {{\bf{h}}_k^{\rm{H}}{{\bf{p}}_k}} \right\} > 0,\;{\mathop{\rm Im}\nolimits} \left\{ {{\bf{h}}_k^{\rm{H}}{{\bf{p}}_k}} \right\} = 0,\;\forall k \in {\cal K}.
\end{align}
Hence, the constraint in \eqref{p_gamma} can be written as
\begin{align}\label{Re1}
    {\mathop{\rm Re}\nolimits} \left\{ {{\bf{h}}_k^{\rm{H}}{{\bf{p}}_k}} \right\} \ge \sqrt {{\gamma _k}} \left\| {{\bf{h}}_k^{\rm{H}}{{\bf{\Delta }}_{{\cal K}\backslash \left\{ k \right\}}},{\sigma _k}} \right\|,\;\forall k \in {\cal K},
\end{align}
where ${{\bf{\Delta }}_{{\cal K}\backslash \left\{ k \right\}}} \buildrel \Delta \over = \left[ {{{\bf{p}}_1}, \ldots ,{{\bf{p}}_{k - 1}},{{\bf{p}}_{k + 1}}, \ldots ,{{\bf{p}}_K}} \right] \in \mathbb{C}{^{L \times \left( {K - 1} \right)}}$. Then, we utilize the inequality $2xy \le {\left( {ax} \right)^2} + {\left( {{y \mathord{\left/
 {\vphantom {y a}} \right.
 \kern-\nulldelimiterspace} a}} \right)^2}$ for $\forall x,y \ge 0$, and the non-convex constraint is approximated by the following convex constraint using a sequential convex programming (SCP) method, specifically SCA, where the constraint is convexified iteratively at each step as
 \begin{align}\label{Re2}
 2{\mathop{\rm Re}\nolimits} \left\{ {{\bf{h}}_k^{\rm{H}}{{\bf{p}}_k}} \right\} \!\ge\! \frac{{{\gamma _k}}}{{q_k^{\left( i \right)}}} \!+\! q_k^{\left( i \right)}\left\| {{\bf{h}}_k^{\rm{H}}{{\bf{\Delta }}_{{\cal K}\backslash \left\{ k \right\}}},{\sigma _k}} \right\|,\;\forall k \in {\cal K},
\end{align}
where $q_k^{\left( i \right)}$ is given by
\begin{align}
    q_k^{\left( i \right)} = {{\sqrt {\gamma _k^{\left( {i - 1} \right)}} } \mathord{\left/
 {\vphantom {{\sqrt {\gamma _k^{\left( {i - 1} \right)}} } {\left\| {{\bf{h}}_k^{\rm{H}}{\bf{\Delta }}_{{\cal K}\backslash \left\{ k \right\}}^{\left( {i - 1} \right)},{\sigma _k}} \right\|}}} \right.
 \kern-\nulldelimiterspace} {\left\| {{\bf{h}}_k^{\rm{H}}{\bf{\Delta }}_{{\cal K}\backslash \left\{ k \right\}}^{\left( {i - 1} \right)},{\sigma _k}} \right\|}}.
\end{align}
The auxiliary variable \(q_k^{(i)}\) plays a pivotal role in the iterative process, dynamically updating the approximation of the interaction terms to ensure a valid lower bound for the original constraint. This iterative convexification method preserves computational tractability and facilitates convergence to a feasible solution that aligns with the original problem structure. By capturing the coupling between the beamforming vectors \({\bf{p}}_k\) and the channel gains \({\bf{h}}_k^{\rm{H}}\), as well as the interference and noise terms \({\bf{\Delta}}_{{\cal K} \backslash \{k\}}\) and \(\sigma_k\), the convexified constraint effectively addresses the non-convexity of the problem.
The superscript $\left(i\right)$ identifies the variables obtained as solutions at the $i$-th iteration.

Note that the power consumption due to the RF chains is determined by the indicator function ${\rm{I}}\left(  \cdot  \right)$, which is not smooth and not differentiable. To address this issue, we reformulate ${\rm{I}}\left(  \cdot  \right)$ as follows
\begin{align}\label{I(x)}
    {\rm{I}}\left( x \right) = \mathop {\lim }\limits_{\varepsilon  \to 0} \frac{{\log \left( {1 + {x \mathord{\left/
 {\vphantom {x \varepsilon }} \right.
 \kern-\nulldelimiterspace} \varepsilon }} \right)}}{{\log \left( {1 + {1 \mathord{\left/
 {\vphantom {1 \varepsilon }} \right.
 \kern-\nulldelimiterspace} \varepsilon }} \right)}},x  \ge  0.
\end{align}
By utilizing small $\varepsilon  > 0$, the indicator function can be approximated by the concave function $g\left( x \right) \buildrel \Delta \over = {{\log \left( {1 + {x \mathord{\left/
 {\vphantom {x \varepsilon }} \right.
 \kern-\nulldelimiterspace} \varepsilon }} \right)} \mathord{\left/
 {\vphantom {{\log \left( {1 + {x \mathord{\left/
 {\vphantom {x \varepsilon }} \right.
 \kern-\nulldelimiterspace} \varepsilon }} \right)} {\log \left( {1 + {1 \mathord{\left/
 {\vphantom {1 \varepsilon }} \right.
 \kern-\nulldelimiterspace} \varepsilon }} \right)}}} \right.
 \kern-\nulldelimiterspace} {\log \left( {1 + {1 \mathord{\left/
 {\vphantom {1 \varepsilon }} \right.
 \kern-\nulldelimiterspace} \varepsilon }} \right)}}$. Furthermore, using the first-order Taylor expansion, $g\left( x \right)$ can be further approximated around the point $\bar x$ by the following expression
 \begin{align}
     \tilde g\left( {x,\bar x} \right) = \frac{{\log \left( {1 + {{\bar x} \mathord{\left/
 {\vphantom {{\bar x} \varepsilon }} \right.
 \kern-\nulldelimiterspace} \varepsilon }} \right) + \frac{1}{{\bar x + \varepsilon }}\left( {x - \bar x} \right)}}{{\log \left( {1 + {1 \mathord{\left/
 {\vphantom {1 \varepsilon }} \right.
 \kern-\nulldelimiterspace} \varepsilon }} \right)}}.
 \end{align}
Finally, through the aforementioned transformations, given $\left\{ {{{\bf{\Phi }}_m},t} \right\}$, the problem ${\cal{P}}^{\rm{BS}}$ in \eqref{P_BS} can be solved iteratively by applying the SCA method. At the $i$-th SCA iteration, the corresponding convex problem to be solved is given by
\begin{subequations}\label{P_BS}
  \begin{align}
&{\cal{\bar P}}^{\rm{BS}}:
\mathop {{\rm{max}}}\limits_{\left\{ {{{\bf{p}}_k},{\gamma _k}} \right\}}  2t\sum\limits_{k = 1}^K {{{\left( {{{\log }_2}\left( {1 + {\gamma _k}} \right)} \right)}^{\frac{1}{2}}}}  - {t^2}\left( {\sum\limits_{k = 1}^K {{{{{\left\| {{{\bf{p}}_k}} \right\|}^2}} \mathord{\left/
 {\vphantom {{{{\left\| {{{\bf{p}}_k}} \right\|}^2}} \eta }} \right.
 \kern-\nulldelimiterspace} \eta }} } \right.\notag\\
&\left. { + P_{{\rm{RF}}}^{{\rm{active}}}\sum\limits_{l = 1}^L {\tilde g\left( {\sum\limits_{k = 1}^K {{{\left| {{\bf{e}}_l^{\rm{T}}{{\bf{p}}_k}} \right|}^2}} ,\sum\limits_{k = 1}^K {{{\left| {{\bf{e}}_l^{\rm{T}}{\bf{p}}_k^{\left( {i - 1} \right)}} \right|}^2}} } \right)} } \right)\\
&\quad\quad\quad\quad{\rm{s}}{\rm{.t}}{\rm{.}}\quad \;\;
\eqref{0-3},\eqref{0-4},\eqref{1-2},\eqref{Re-Im},\eqref{Re2}. \notag
\end{align}
\end{subequations}
It is important to note that during the optimization process, it is necessary to ensure that the number of selected antennas remains greater than or equal to the number of UEs $K$ at all times. This constraint guarantees the feasibility and effectiveness of multi-user communication. By iteratively solving the formulated convex optimization problem ${\cal{\bar P}}^{\rm{BS}}$, the algorithm gradually refines the solution, ultimately yielding the optimized digital precoding matrices ${\bf{p}}_k$.

\begin{rem}
Unlike heuristic antenna selection methods such as threshold-based schemes \cite{sanayei2004antenna,shi2024joint}, which rely on predefined criteria, the proposed joint design dynamically optimizes both antenna selection and beamforming. This enables better adaptability to channel conditions, prevents suboptimal BS activation, and enhances the EE by considering the interplay between BS antenna selection and beamforming in a unified optimization framework.
\end{rem}

\subsection{Optimization of $\left\{ {{{\bf{\Phi }}_m}} \right\}$ Given $\left\{ {{{\bf{p}}_k},t} \right\}$ }\label{opt_phase}
In this section, we propose two different methods, i.e., an SDP-based method and ab iterative gradient ascent method, to optimize the wave-based beamforming ${\left\{ {{{\bf{\Phi }}_m}} \right\}}$ of the SIM given $\left\{ {{{\bf{p}}_k},t} \right\}$.

\subsubsection{SDP-based Method}
The SDP method provides a convex relaxation approach for solving non-convex optimization problems, offering near-optimal solutions with high accuracy, making it highly effective for optimizing communication systems \cite{vandenberghe1996semidefinite}. Therefore, we propose an SDP-based layer-by-layer iterative optimization algorithm for SIM-based beamforming.
First, to facilitate subsequent calculations, we reformulate \eqref{gamma_k} as follows:
\begin{align}
    {\gamma _k} = \frac{{{{\left| {{\bf{\varphi }}_m^{\rm{H}}{\bf{H}}_{k,m}^{\rm{H}}{{{\bf{\bar w}}}_k}} \right|}^2}}}{{\mathop \sum \limits_{j = 1,j \ne k}^K {{\left| {{\bf{\varphi }}_m^{\rm{H}}{\bf{H}}_{k,m}^{\rm{H}}{{{\bf{\bar w}}}_j}} \right|}^2} + \sigma _k^2}},
\end{align}
where ${\bf{\varphi }}_m^{\rm{H}} = {\rm{diag}}{\left( {{{\bf{\Phi }}_m}} \right)^{\rm{T}}} = \left[ {{e^{j\varphi _m^1}}, \ldots ,{e^{j\varphi _m^N}}} \right] \in \mathbb{C}{^{1 \times N}}$, ${{{\bf{\bar w}}}_k} = {{\bf{W}}_1}{{\bf{p}}_k}$, and ${\bf{H}}_{k,m}^{\rm{H}} \in \mathbb{C}{^{N \times N}}$ is expressed in \eqref{H_km} shown at the top of this page.
\begin{figure*}[t!]
\normalsize
\setcounter{mytempeqncnt}{1}
\setcounter{equation}{30}
\begin{align}\label{H_km}
{\bf{H}}_{k,m}^{\rm{H}} \buildrel \Delta \over = {\rm{diag}}\left( {{\bf{h}}_{{\rm{SIM}},k}^{\rm{H}}{{\bf{\Phi }}_M}{{\bf{W}}_M} \ldots {{\bf{\Phi }}_{m + 1}}{{\bf{W}}_{m + 1}}} \right){{\bf{W}}_m}{{\bf{\Phi }}_{m - 1}}{{\bf{W}}_{m - 1}} \ldots {{\bf{\Phi }}_2}{{\bf{W}}_2}{{\bf{\Phi }}_1}{{\bf{W}}_1}.
\end{align}
\setcounter{equation}{31}
\hrulefill
\end{figure*}
Then, by defining the matrices
\begin{align}
&{{\bf{V}}_m} \buildrel \Delta \over = {\left[ {{\bf{\varphi }}_m^{\rm{H}},1} \right]^{\rm{H}}}\left[ {{\bf{\varphi }}_m^{\rm{H}},1} \right],\\
&{\bf{U}}_{kj}^m \buildrel \Delta \over = {\left[ {{{\left( {{\bf{H}}_{k,m}^{\rm{H}}{{{\bf{\bar w}}}_k}} \right)}^{\rm{H}}},0} \right]^{\rm{H}}}\left[ {{{\left( {{\bf{H}}_{k,m}^{\rm{H}}{{{\bf{\bar w}}}_j}} \right)}^{\rm{H}}},0} \right],
\end{align}
we obtain
\begin{align}
{\left| {{\bf{h}}_{{\rm{SIM}},k}^{\rm{H}}{\bf{G}}{{\bf{W}}_1}{{\bf{p}}_j}} \right|^2} = {\rm{Tr}}\left( {{{\bf{V}}_m}{\bf{U}}_{kj}^m} \right).
\end{align}
Therefore, given $\left\{ {{{\bf{p}}_k},t} \right\}$, the SIM optimization problem in \eqref{P_0} is equivalent to problem ${\cal{P}}^{\rm{SDP}}$ as follows
\begin{subequations}\label{P_SDP}
\begin{align}
&{\cal{P}}^{\rm{SDP}}:\mathop {{\rm{max}}}\limits_{\left\{ {{{\bf{V}}_m},{\gamma _k}} \right\}} \quad \sum\limits_{k = 1}^K {\log } \left( {1 + {\gamma _k}} \right)\label{P_SDP1}\\
&{\rm{s}}{\rm{.t}}{\rm{.}} \;\;{\rm{Tr}}\left( {{{\bf{V}}_m}{\bf{U}}_{kk}^m} \right) \!\ge \!{\gamma _k}\! {\left(\sum\limits_{j \in {\cal K}\backslash \left\{ k \right\}} \!\!\!\!\!{{\rm{Tr}}\!\left( {{{\bf{V}}_m}{\bf{U}}_{kj}^m} \right) \!+\! \sigma _k^2} \right)} ,\!\forall k \!\in\! {\cal K},\label{P_SDP2}\\
&\quad \quad \;\;{\gamma _k} \ge \gamma _k^{\min },\forall k \in {\cal K},\label{P_SDP3}\\
&\quad \quad \;\;{\left[ {{{\bf{V}}_m}} \right]_{n,n}} = 1,\forall m \in {\cal M},n = 1, \ldots ,N + 1,\label{P_SDP4}\\
&\quad \quad \;\;{{\bf{V}}_m}\succeq{\bf{0}},{\rm{rank}}\left( {{{\bf{V}}_m}} \right) = 1,\forall m \in {\cal M},\label{P_SDP5}
\end{align}
\end{subequations}
However, due to the coupling of variables coupling in the constraint \eqref{P_SDP2} and the rank-one constraint on ${{{\bf{V}}_m}}$, the problem in \eqref{P_SDP} is not convex. Therefore, we introduce the auxiliary variables $\left\{ {{\mu _k}} \right\}$ and re-express the non-convex constraint in \eqref{P_SDP2} as
\begin{align}
&{\rm{Tr}}\left( {{{\bf{V}}_m}{\bf{U}}_{kk}^m} \right) \ge \mu _k^2,\forall k \in {\cal K},\label{TRVS}\\
&{{\mu _k^2} \mathord{\left/
 {\vphantom {{\mu _k^2} {{\gamma _k}}}} \right.
 \kern-\nulldelimiterspace} {{\gamma _k}}} \ge \sum\limits_{j \in {\cal K}\backslash \left\{ k \right\}} {{\rm{Tr}}\left( {{{\bf{V}}_m}{\bf{U}}_{kj}^m} \right) + \sigma _k^2} ,\forall k \in {\cal K}.\label{mu/gamma}
\end{align}
The left-hand side of the constraint in \eqref{mu/gamma} is jointly convex with respect to $\left\{ {{\mu _k}} \right\}$ and $\left\{ {{\gamma _k}} \right\}$. By applying the first-order Taylor expansion, an approximation of the constraint in \eqref{mu/gamma} at the point $\left( {\mu _k^{\left( \ell  \right)},\gamma _k^{\left( \ell  \right)}} \right)$ can be derived as follows:
\begin{align}\label{2mumu/gamma}
    \frac{{2\mu _k^{\left( \ell  \right)}{\mu _k}}}{{\gamma _k^{\left( \ell  \right)}}} - \frac{{{{\left( {\mu _k^{\left( \ell  \right)}} \right)}^2}{\gamma _k}}}{{{{\left( {\gamma _k^{\left( \ell  \right)}} \right)}^2}}} \ge \!\!\!\sum\limits_{j \in {\cal K}\backslash \left\{ k \right\}} \!\!\!\!{{\rm{Tr}}\left( {{{\bf{V}}_m}{\bf{U}}_{kj}^m} \right) + \sigma _k^2} ,\forall k \in {\cal K}.
\end{align}

Then, for the rank-one constraint \eqref{P_SDP5}, we have
\begin{align}
    {\lambda _{\max }}\left( {{{\bf{V}}_m}} \right) = {\rm{Tr}}\left( {{{\bf{V}}_m}} \right),
\end{align}
where ${\lambda _{\max }}\left( {{{\bf{V}}_m}} \right)$ denotes the largest eigenvalue of ${{{\bf{V}}_m}}$. To proceed, we relax the rank-one constraint on ${{{\bf{V}}_m}}$ by \cite{mu2020exploiting}
\begin{align}\label{lambda_max}
    {\lambda _{\max }}\left( {{{\bf{V}}_m}} \right) \ge {\varepsilon _m}{\rm{Tr}}\left( {{{\bf{V}}_m}} \right),
\end{align}
where ${\varepsilon _m} \in \left[ {0,1} \right]$ is a the relaxation parameter. Specifically, if the equality holds for ${\varepsilon _m} = 1$, we have ${\rm{rank}}\left( {{{\bf{V}}_m}} \right) = 1$. If ${\varepsilon _m} = 0$,  the rank-one constraint is dropped. Consequently, we can incrementally increase ${\varepsilon _m}$ from 0 to 1, ultimately arriving at a rank-one solution.
Furthermore, for the non-differentiable function ${\lambda _{\max }}\left( {{{\bf{V}}_m}} \right)$, we use the formulation ${\lambda _{\max }}\left( {{{\bf{V}}_m}} \right) = {\max _{{\bf{x}} \in \mathbb{C}{^{N + 1}},{{\left\| {\bf{x}} \right\|}^2} \le 1}}{{\bf{x}}^{\rm{H}}}{{\bf{V}}_m}{\bf{x}}$ to approximate \eqref{lambda_max} as
\begin{align}\label{sigma_max}
    {{\bf{\zeta }}_{\max }}{\left( {{\bf{V}}_m^{\left( \ell  \right)}} \right)^{\rm{H}}}{{\bf{V}}_m}{{\bf{\zeta }}_{\max }}\left( {{\bf{V}}_m^{\left( \ell  \right)}} \right) \ge \varepsilon _m^{\left( \ell  \right)}{\rm{Tr}}\left( {{{\bf{V}}_m}} \right),
\end{align}
where ${{\bf{\zeta }}_{\max }}\left( {{\bf{V}}_m^{\left( \ell  \right)}} \right)$ denotes the eigenvector corresponding to ${\lambda _{\max }}\left( {{{\bf{V}}_m}} \right)$.

Finally, the problem ${\cal{P}}^{\rm{SDP}}$ in \eqref{P_SDP} can be tackled iteratively by solving the following convex problem at the $\left( {\ell  + 1} \right)$-th iteration
\begin{subequations}\label{P_SDP}
\begin{align}
{\cal{\hat P}}^{\rm{SDP}}: &\mathop {{\rm{max}}}\limits_{\left\{ {{{\bf{V}}_m},{\gamma _k},{\mu _k}} \right\}} \quad \sum\limits_{k = 1}^K {\log } \left( {1 + {\gamma _k}} \right)\\
&\quad\quad{\rm{s}}{\rm{.t}}{\rm{.}}\quad \;\;\;\;\;{{\bf{V}}_m} \succeq{\bf{0}},\forall m \in {\cal M},\\
&\quad\quad\quad \quad \;\;\;\;\;\;\,\eqref{P_SDP3}, \eqref{P_SDP4}, \eqref{TRVS}, \eqref{2mumu/gamma}, \eqref{sigma_max}.\notag
\end{align}
\end{subequations}

\begin{rem}
    The proposed SDP-based iterative optimization algorithm tackles the non-convex SIM wave-based beamforming problem by relaxing the original non-convex constraints into a solvable form. However, the computational complexity of SDP in each SIM layer is generally of the order of $N^3$, which is relatively high, especially when the number of SIM meta-atoms is large.
\end{rem}

\begin{algorithm}[t]
\caption{Proposed AO-based hybrid precoding algorithm for solving the problem in \eqref{P_0} }
\begin{algorithmic}[1]
\State \textbf{Input:} $\mathbf{h}_{{\rm{SIM}},k}$, $\mathbf{W}_{m}$.
\State \textbf{Initialize:} $\varphi _{m}^n \in \left[ {0,2\pi } \right)$ and $\{{\bf{p}}_k\}$ feasible to the problem \eqref{P_0}.
\State Set $r = 0$ and ${\rm{Opt}}^{(r)} = 0$.
\State \textbf{Repeat}
    \State \hspace{\algorithmicindent}Update $r = r+1$.
    \State \hspace{\algorithmicindent}Set $i = 0$, ${\rm{Opt}}_1^{(0)} = 0$ and compute $P_{\rm{total}}^{(r-1)}$ using
    \Statex \hspace{\algorithmicindent}$\{{\bf{p}}_k\}$.
    \State \hspace{\algorithmicindent}\textbf{Repeat}
    \State \hspace{\algorithmicindent}\hspace{\algorithmicindent}Update $i = i+1$.
    \State\hspace{\algorithmicindent}\hspace{\algorithmicindent}Update $t^{(i)}$ according to \eqref{t}.
    \State\hspace{\algorithmicindent}\hspace{\algorithmicindent}Compute $\{{\bf{p}}_k\}$ by solving the problem in \eqref{P_1} and \Statex \hspace{\algorithmicindent}\hspace{\algorithmicindent}update the objective function ${\rm{Opt}}_{1}^{(i)}$.
    \State\hspace{\algorithmicindent}\textbf{Until} ${{\left| {{\rm{Opt}}_1^{\left( i \right)} - {\rm{Opt}}_1^{\left( {i - 1} \right)}} \right|} \mathord{\left/
 {\vphantom {{\left| {{\rm{Opt}}_1^{\left( i \right)} - {\rm{Opt}}_1^{\left( {i - 1} \right)}} \right|} {{\rm{Opt}}_1^{\left( i \right)}}}} \right.
 \kern-\nulldelimiterspace} {{\rm{Opt}}_1^{\left( i \right)}}} \le \varepsilon $
    \State \hspace{\algorithmicindent}Update ${\bf{p}}_k = {\bf{p}}_k^{\left( i \right)}$ and $P_{\rm{total}}^{(r)}$.
\State \hspace{\algorithmicindent}Set $\ell  = 0$ and ${{\rm{Opt}}_2^{\left( 0 \right)}} = 0$.
    \State \hspace{\algorithmicindent}\textbf{Repeat}
    \State \hspace{\algorithmicindent}\hspace{\algorithmicindent}Update $m = m +1$.
         \State \hspace{\algorithmicindent}\hspace{\algorithmicindent}\textbf{Repeat}
           \State \hspace{\algorithmicindent}\hspace{\algorithmicindent}\hspace{\algorithmicindent}Update $\ell = \ell +1$.
           \State \hspace{\algorithmicindent}\hspace{\algorithmicindent}\hspace{\algorithmicindent}Compute $\{{{{\bf{\Phi }}_m}}\}$ by solving the SDP-based
           \Statex \hspace{\algorithmicindent}\hspace{\algorithmicindent}\hspace{\algorithmicindent}problem in \eqref{P_SDP} or the PGA-based problem in
           \Statex \hspace{\algorithmicindent}\hspace{\algorithmicindent}\hspace{\algorithmicindent}\eqref{update_phi}. Update the objective function ${{\rm{Opt}}_2^{\left( \ell \right)}}$.
           \State \hspace{\algorithmicindent}\hspace{\algorithmicindent}\textbf{Until} ${{\left| {{\rm{Opt}}_2^{\left( \ell  \right)} - {\rm{Opt}}_2^{\left( {\ell  - 1} \right)}} \right|} \mathord{\left/
 {\vphantom {{\left| {{\rm{Opt}}_2^{\left( \ell  \right)} - {\rm{Opt}}_2^{\left( {\ell  - 1} \right)}} \right|} {{\rm{Opt}}_2^{\left( \ell  \right)}}}} \right.
 \kern-\nulldelimiterspace} {{\rm{Opt}}_2^{\left( \ell  \right)}}} \le \varepsilon $.
          \State \hspace{\algorithmicindent}\textbf{Until} $m = M$.
\State \textbf{Until} ${{\left| {{\rm{Op}}{{\rm{t}}^{\left( r \right)}} - {\rm{Op}}{{\rm{t}}^{\left( r \right)}}} \right|} \mathord{\left/
 {\vphantom {{\left| {{\rm{Op}}{{\rm{t}}^{\left( r \right)}} - {\rm{Op}}{{\rm{t}}^{\left( r \right)}}} \right|} {{\rm{Op}}{{\rm{t}}^{\left( r \right)}}}}} \right.
 \kern-\nulldelimiterspace} {{\rm{Op}}{{\rm{t}}^{\left( r \right)}}}} \le \varepsilon $.
\State \textbf{Output:} The BS digital precoding vector $\{{\bf{p}}_{k}\}$, SIM wave-based beamforming matrices $\{{{{\bf{\Phi }}_m}}\}$, and $\rm{EE}$.
\end{algorithmic}
\end{algorithm}

\subsubsection{PGA-based Method}
In this section, we introduce a computationally efficient PGA algorithm to update the phase shifts ${\varphi _m^n}$ iteratively until converging to the vicinity of a stationary point. The detailed steps of the PGA algorithm are as follows:
\begin{enumerate}
\item[\textit{i)}] First, we initialize the phase shifts of all SIM meta-atoms $\varphi _{m}^n \in [0,2\pi ), \forall m \in {\cal M},\forall n \in {\cal N}$. Next, we compute the sum rate by using \eqref{R_sum}.
\item[\textit{ii)}] Then, we derive the partial derivative of $R_{\rm{sum}}$ with respect to all $\varphi _{m}^n$, as given in Theorem \ref{them}.
\begin{them}\label{them}
For $\forall m \in {\cal M},\forall n \in {\cal N}$, the partial derivative of the sum rate $R_{\rm{sum}}$ with respect to $\varphi _{m}^n$ is given by
\begin{align}\label{deta_R}
\setcounter{equation}{42}
\frac{{\partial {R_{{\rm{sum}}}}}}{{\partial \varphi _m^n}} \!=\! {\log _2}e \sum\limits_{k = 1}^K {{\chi _k}\!\left(\! {{{\left( {\varpi _m^n} \right)}_{k,k}} \!-\! {\gamma _k}\!\!\!\!\sum\limits_{j = 1,j \ne k}^K \!\!\!{{{\left( {\varpi _m^n} \right)}_{k,j}}} } \right)},
\end{align}
where ${\chi _k}$ is defined as
\begin{align}\label{dalta}
  {\chi _k} = \frac{1}{{\sum\limits_{j = 1}^K {{{\left| {{\bf{h}}_{{\rm{SIM}},k}^{\rm{H}}{\bf{G}}{{\bf{W}}_1}{{\bf{p}}_j}} \right|}^2} + \sigma _k^2} }},
\end{align}
and ${\left( {\varpi _m^n} \right)_{k,j}}$ is defined in \eqref{varpi} show at the top of the next page. Furthermore, in \eqref{varpi}, ${{\bf{b}}_m^n}$ and ${{{\left( {{\bf{q}}_{m}^n} \right)}^{\rm{H}}}}$ denote the $n$-th column of matrix ${{\bf{B}}_{m}} \in {\mathbb{C}^{N \times N}}$ and $n$-th row of matrix ${{\bf{Q}}_{m}} \in {\mathbb{C}^{N \times N}}$, respectively. Specially, the matrices ${{\bf{B}}_{m}}$ and ${{\bf{Q}}_{m}}$ are defined as follows:
\begin{figure*}[t!]
\normalsize
\setcounter{mytempeqncnt}{1}
\setcounter{equation}{44}
\begin{align}\label{varpi}
{\left( {\varpi _m^n} \right)_{k,j}} = \frac{{\partial {{\left| {{\bf{h}}_{{\rm{SIM}},k}^{\rm{H}}{\bf{G}}{{\bf{W}}_1}{{\bf{p}}_k}} \right|}^2}}}{{\partial \varphi _m^n}} = 2 \Im \left[ {\left( {{e^{j\varphi _m^n}}{\bf{h}}_{{\rm{SIM}},k}^{\rm{H}}{\bf{b}}_m^n{{\left( {{\bf{q}}_m^n} \right)}^{\rm{H}}}{{\bf{W}}_1}{{\bf{p}}_k}} \right)\left( {{\bf{h}}_{{\rm{SIM}},k}^{\rm{H}}{\bf{G}}{{\bf{W}}_1}{{\bf{p}}_k}} \right)} \right].
\end{align}
\setcounter{equation}{45}
\hrulefill
\end{figure*}
\begin{align}\label{B}
    {{\bf{B}}_{m}} \!\!\buildrel \Delta \over =\!\! \left\{ \!\!\!{\begin{array}{*{20}{c}}
{{{\bf{\Phi }}_{M}}{{\bf{W}}_{M}} \!\ldots\! {{\bf{\Phi }}_{\left( {m + 1} \right)}}{\!{\bf{W}}_{\left( {m + 1} \right)}},}&\!\!\!\!\!{{\rm{if}}\,m \!\ne\! M,}\\
{{{\bf{I}}_N},}&\!\!\!\!\!{{\rm{if}}\,m \!=\! M,}
\end{array}} \right.
\end{align}
\begin{align}\label{Q}
{{\bf{Q}}_{m}} \buildrel \Delta \over = \left\{ {\begin{array}{*{20}{c}}
{{{\bf{W}}_{m}}{{\bf{\Phi }}_{\left( {m - 1} \right)}} \ldots {{\bf{W}}_{2}}{{\bf{\Phi }}_{1}},}&{{\rm{if}}\,m \ne 1,}\\
{{{\bf{I}}_N},}&{{\rm{if}}\,m = 1,}
\end{array}} \right.
\end{align}
\end{them}

\begin{IEEEproof}
    The proof is given in Appendix A.
\end{IEEEproof}

\item[\textit{iii)}] Based on the partial derivatives of $R_{\rm{sum}}$ with respect to ${\varphi_{m}^n}$ given in \eqref{deta_R}, we subsequently update all the phase shifts ${\varphi_{m}^n}$ simultaneously as follows:
\begin{align}\label{update_phi}
    \varphi _{m}^n \leftarrow \varphi _{m}^n + \xi \frac{{\partial {R_{{\rm{sum}}}}}}{{\partial \varphi _{m}^n}},\forall m \in {\cal M},\forall n \in {\cal N},
\end{align}
where $\xi  > 0$ denotes the $\mathit{Armijo}$ step size, which is determined using the backtracking line search method \cite{papazafeiropoulos2021intelligent}.

\item[\textit{iv)}] Finally, we repeat steps \textit{ii)} to \textit{iii)} iteratively until the fractional increase in the sum rate is less than a predetermined threshold. Eventually, we obtain the values of all $\varphi _{m}^n$, which are the optimized phase shifts of the SIM.
\end{enumerate}

\begin{rem}
    The convergence of the PGA algorithm to a local maximum is guaranteed due to two key factors: (1) the sum rate $R_{\rm{sum}}$ is upper bounded as demonstrated in \cite[Eq.~(17)]{an2023stacked2}, and (2) the sum rate $R_{\rm{sum}}$ exhibits a non-decreasing behavior when an appropriate Armijo step size $\xi$ is chosen for each iteration \cite{armijo1966minimization}.
    To enhance the robustness of the PGA algorithm, multiple initializations are employed.
\end{rem}

\subsection{Complexity Analysis}
The complexity of the proposed framework consists of two main parts: the complexity associated with the BS digital precoding vectors ${\bf{p}}_k$ in \eqref{P_BS} and the complexity of SIM wave-based beamforming matrices $\left\{ {{{\bf{\Phi }}_m}} \right\}$ in \eqref{P_SDP} or \eqref{update_phi}.
For solving the problem in \eqref{P_BS}, the computational complexity of each iteration is ${\cal{O}}_{\rm{BS}} = {\cal{O}}(L^3)$, by using second-oder cone programming.
For solving the problem in \eqref{P_SDP}, the computational complexity of each iteration and for each SIM layer is ${\cal{O}}(N^{4.5})$ by using semidefinite programming. Therefore, the computational complexity for optimizing all the meta-atoms of the SIM is ${\cal{O}}_{\rm{SDP}} = {\cal{O}}(MN^{4.5})$.
By adopting the PGA-based algorithm, the computational complexity is given by $\mathcal{O}_{\rm{PGA}} = \mathcal{O}(MNI_{\rm{PGA}}(4N+3))$, where $I_{\rm{PGA}}$ denotes the number of iterations for achieving convergence.
As a result, when the SDP-based method is applied, the total complexity of the proposed AO-based hybrid precoding algorithm can be expressed as $\mathcal{O}_{\rm{total}} = I_{\rm{AO}}({\cal{O}}(I_{\rm{BS}}L^3)+{\cal{O}}({I_{\rm{SDP}}}MN^{4.5}))$, where $I_{\rm{AO}}$, $I_{\rm{BS}}$ and ${I_{\rm{SDP}}}$ denote the number of iterations for AO, for solving the BS digital precoding problem in \eqref{P_BS}, and for solving the SIM wave-based beamforming in \eqref{P_SDP}, respectively. When the PGA-based algorithm is applied, the total computational complexity is given by ${{\cal O}_{{\rm{total}}}} = {I_{{\rm{AO}}}}\left( {{\cal O}\left( {{I_{{\rm{BS}}}}{L^3}} \right) + {\cal O}\left( {{I_{{\rm{PGA}}}}MN\left( {N + 3} \right)} \right)} \right)$.
Therefore, the computational complexity of the PGA-based algorithm is much lower than that of the SDP-based algorithm, especially when the number of SIM meta-atoms $N$ is sufficiently large.
\begin{table*}[t]
\centering
    \fontsize{8}{12}\selectfont
    \caption{Simulation Parameters.}
    \label{Table1}
    \begin{tabular}{cc|cc}
    \toprule
    \bf Parameters &  \bf Values & \bf Parameters &  \bf Values \\
    \midrule
    Noise power $\sigma^2$ & $-174\, \text{dBm/Hz}$ & Fixed power consumption of the SIM controller $P_{\rm{controller}}$ & $25\, \text{dBm}$ \\
    Transmission bandwidth \cite{an2023stacked2} & $10\, \text{MHz}$ & Fixed power consumption of each UE $P_{\rm{UE}}$ & $20\; \rm{dBm}$  \\
    Static power consumption of the BS $P_{\rm{S}}$ & $4.5\, \text{W}$ & Power consumption of each meta-atom $P_{\rm{meta}}$ & $10\; \rm{dBm}$  \\
    RF chain power consumption $P_{{\rm{RF}}}^{{\rm{active}}}$ \cite{tervo2018energy} & $0.4\,\rm{W}$ & Maximum transmit power of each antenna $P_{{l}}^{\rm{max}}$ & $30\; \rm{dBm}$  \\
    \bottomrule
    \end{tabular}
\end{table*}
\section{Numerical Results}
\subsection{Simulation Setup}
This section presents numerical results to analyze the proposed algorithms and evaluate the performance of the considered SIM-based multi-user MISO system. The simulation setting is given as follows. A BS serves $K$ single-antenna UEs, and we assume that the distance between the BS and the UE cluster is $100\,\text{m}$, while each UE experiences a random movement within a range of $5\,\text{m}$.
Furthermore, an SIM, which stacks multiple metasurface layers, is integrated into the BS to execute transmit beamforming in the wave domain. The height of the SIM-assisted BS and UEs is 15 $\rm{m}$ and 1.65 $\rm{m}$, respectively. Furthermore, we assume that the thickness of the SIM is $T_{\rm{SIM}} = 5\lambda$, ensuring that the spacing between two adjacent metasurfaces for an $M$-layer SIM is ${d_{{\rm{Layer}}}} = {{{T_{{\rm{SIM}}}}} \mathord{\left/
 {\vphantom {{{T_{{\rm{SIM}}}}} M}} \right.
 \kern-\nulldelimiterspace} M}$.
Furthermore, we consider a square metasurface arrangement with $N={N_x}{N_y}$ meta-atoms where $N_x = N_y$, and $N_x$ and $N_y$ denote the number of meta-atoms along the $x$-axis and $y$-axis, respectively. Furthermore, we assume half-wavelength spacing between adjacent antennas/meta-atoms at the BS and metasurface layers. Also, the size of each meta-atom is $d = {d_x} = {d_y} = {\lambda  \mathord{\left/
 {\vphantom {\lambda  2}} \right.
 \kern-\nulldelimiterspace} 2}$. The propagation coefficients $w^{n,n'}_{m}$ are computed from \eqref{w}.
 We assume a correlated Rayleigh fading channel model, and the distance-dependent path loss is modeled as
 \begin{align}
      {\beta _{k}} = {C_0}{\left( {{{{d_{k}}} \mathord{\left/
 {\vphantom {{{d_{lk}}} {{d_0}}}} \right.
 \kern-\nulldelimiterspace} {{d_0}}}} \right)^{\varpi_{0} }},\quad {d_{k}} > {d_0},
 \end{align}
where $d_{k}$ denotes the link distance between the SIM-assisted BS to the $k$-th UE. Also, ${C_0} = {\left( {{\lambda  \mathord{\left/
 {\vphantom {\lambda  {4\pi {d_0}}}} \right.
 \kern-\nulldelimiterspace} {4\pi {d_0}}}} \right)^2}$ denotes the free space path loss with respect to the reference distance $d_0 = 1$ m \cite{rappaport2015wideband}, and ${\varpi_{0}} =3.5 $ denotes the path loss exponent. Besides, we consider a system operating at a carrier frequency of 28 GHz with a transmission bandwidth of 10 MHz, and an effective noise power spectral density of $-174$ dBm/Hz.
The RF chain power consumption $P_{{\rm{RF}}}^{{\rm{active}}}$ is $0.4\,\rm{W}$, and the static power consumption of the BS $P_{\rm{S}}$ is $4.5\,\rm{W}$ \cite{tervo2018energy}. The fixed power consumption of the SIM controller $P_{\rm{controller}}$ is $25\,\rm{dBm}$.\footnote{
As mentioned in \cite{liu2022programmable}, an SIM is a semi-passive structure with low power consumption. Currently, there is no large-scale experimental data quantifying the exact power consumption of SIM controllers. Since $P_{\rm{controller}}$ is a fixed value in our model, it does not affect the overall trends and conclusions, but a precise quantification of it remains an important topic for future research.}
Each UE transmits with maximum power $20\; \rm{mW}$. We assume that the activated threshold of the RF chain is ${\rho _l} = 10^{-4}$. The key simulation parameters are summarized in Table \ref{Table1}. As for the PGA algorithm, the maximum number of iterations is set to 100, and the initial learning rate and decay parameter are set to 0.1 and 0.5, respectively, unless specified otherwise. The simulation results are obtained by averaging over 100 independent experiments. For comparison, the following schemes are considered.

\subsubsection{\textbf{Hybrid-SDP}}
The BS adopts digital precoding, and the SDP algorithm is adopted for SIM wave-based beamforming.
\subsubsection{\textbf{Hybrid-PGA}}
The BS adopts digital precoding, and the PGA algorithm is adopted for SIM wave-based beamforming.
\subsubsection{\textbf{Digital-Pre}}
The BS adopts digital precoding, and SIM adopts random phase shifts.
\subsubsection{\textbf{Wave-SIM}}
The BS does not adopt digital precoding (each antenna transmits a single stream), but only SIM wave-based beamforming is utilized.

\begin{figure}[t]
\centering
\includegraphics[scale=0.5]{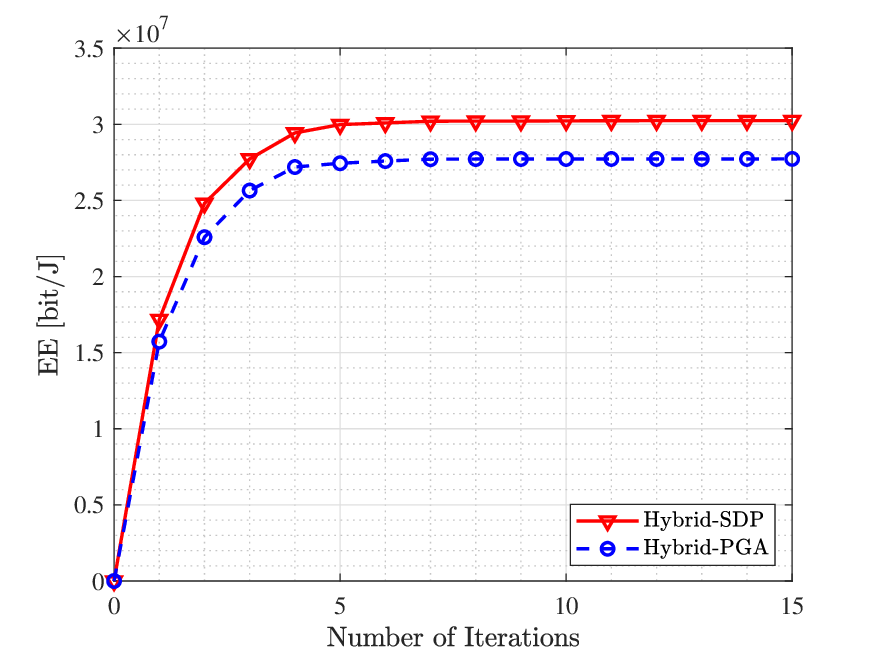}
\caption{EE against the number of iterations ($L = 4$, $M = 4$, $K = 4$, ${{P}_{\rm{max}}} = 35 \,\rm{dBm}$, ${{{P}^{\rm{max}}_{l}}} = 30 \,\rm{dBm}$, ${{P}_{\rm{meta}}} = 10 \,\rm{dBm}$, ${{{\gamma}^{\rm{min}}_{k}}} = 0 \,\rm{dB}$).}\label{Fig_iteration}\vspace{-0.2cm}
\end{figure}
As shown in Fig.~\ref{Fig_iteration}, the proposed Hybrid-SDP and PGA algorithms can achieve convergence within a few iterations. Also, after achieving convergence, the Hybrid-SDP algorithm enhances the EE performance by 7\%, underscoring the effectiveness of the proposed SDP algorithm.

\begin{figure}[t]
\centering
\includegraphics[scale=0.5]{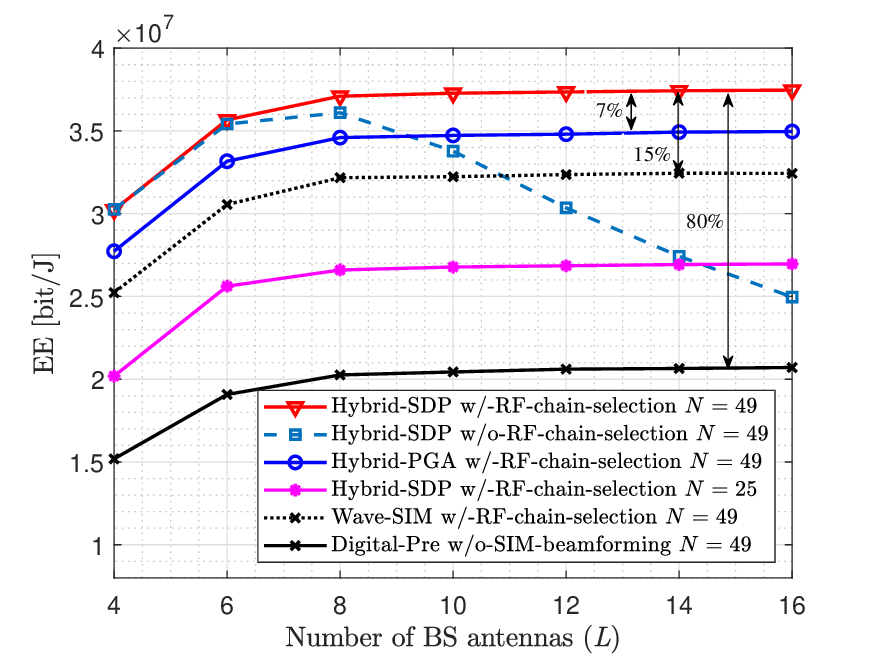}
\caption{Average EE against the number of BS antennas and the number of SIM meta-atoms per layer considering the SDP/PGA algorithms ($M = 4$, $K = 4$, ${{P}_{\rm{max}}} = 35 \,\rm{dBm}$, ${{{P}^{\rm{max}}_{l}}} = 30 \,\rm{dBm}$, ${{P}_{\rm{meta}}} = 10 \,\rm{dBm}$, ${{{\gamma}^{\rm{min}}_{k}}} = 0 \,\rm{dB}$).}\label{Fig_L_EE}\vspace{-0.2cm}
\end{figure}
Fig.~\ref{Fig_L_EE} illustrates the average EE against the number of BS antennas and SIM meta-atoms per layer by considering the hybrid SDP and PGA algorithms. The schemes without RF chain selection (i.e., antenna selection), only SIM wave-based beamforming without digital precoding (Wave-SIM), and only digital precoding without SIM beamforming (Digital-Pre) are also plotted as benchmark schemes. The digital precoding vectors are obtained by applying the SCA method in Section III-C. Without RF chain selection, as the number of BS antennas increases, the system EE initially improves and reaches a maximum value when the number of antennas is $L = 8$. Subsequently, as the number of antennas continues to increase, the EE begins to decline.
This occurs because the power consumption of each RF chain $P_{\rm{RF}}$ is considered. When the number of antennas is $L > 8$, the RF power consumption has a greater impact, dominating the communication capacity gain. However, by utilizing the proposed RF chain selection scheme, the EE continues to increase as the number of BS antennas increases. The reason is that, thanks to the RF chain selection strategy, the antennas with poor channel conditions are deactivated, thereby increasing the EE.
Furthermore, the results indicate that the EE of an SIM-assisted BS adopting the Hybrid-SDP scheme with $L=14$ is enhanced by 7\%, 15\%, and 80\% compared to the Hybrid-PGA scheme, Wave-SIM scheme, and Digital-Pre scheme, respectively. This highlights the advantages of SIM, but an effective beamforming design is required to fully capitalize on its wave-domain processing capabilities.

\begin{figure}[t]
\centering
\includegraphics[scale=0.5]{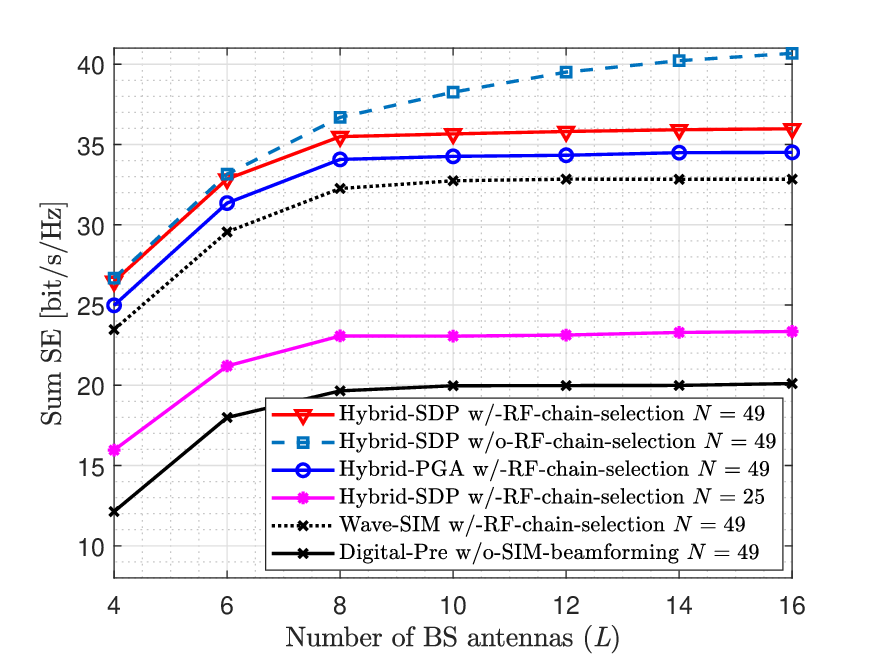}
\caption{Average sum SE against the number of BS antennas and the number of SIM meta-atoms per layer considering the SDP/PGA algorithms ($M = 4$, $K = 4$, ${{P}_{\rm{max}}} = 35 \,\rm{dBm}$, ${{{P}^{\rm{max}}_{l}}} = 30 \,\rm{dBm}$, ${{P}_{\rm{meta}}} = 10 \,\rm{dBm}$, ${{{\gamma}^{\rm{min}}_{k}}} = 0 \,\rm{dB}$).}\label{Fig_L_SE}\vspace{-0.2cm}
\end{figure}
Fig.~\ref{Fig_L_SE} shows the average sum SE against the number of BS antennas and SIM meta-atoms per layer by considering the hybrid SDP and PGA algorithms. When maximizing the EE, the results indicate that the system sum SE progressively increases with the addition of BS antennas, ultimately reaching a plateau.
Specifically, at convergence, the proposed Hybrid-SDP scheme shows a sum SE increase of 10\% and 79\% compared to the Wave-SIM and Digital-Pre scheme, which reflects the advantages of hybrid BS digital and SIM wave-based precoding.
Furthermore, comparing the setups $N=49$ and BS $L=4$, against $N=25$ with $L=8$, we see that the sum SE improves by 10\%.
This indicates that the sum SE can be entrenched by increasing the number of SIM meta-atoms in lieu of increasing the number of BS antennas, hence reducing complexity, cost, and power consumption.

\begin{figure}[t]
\centering
\includegraphics[scale=0.5]{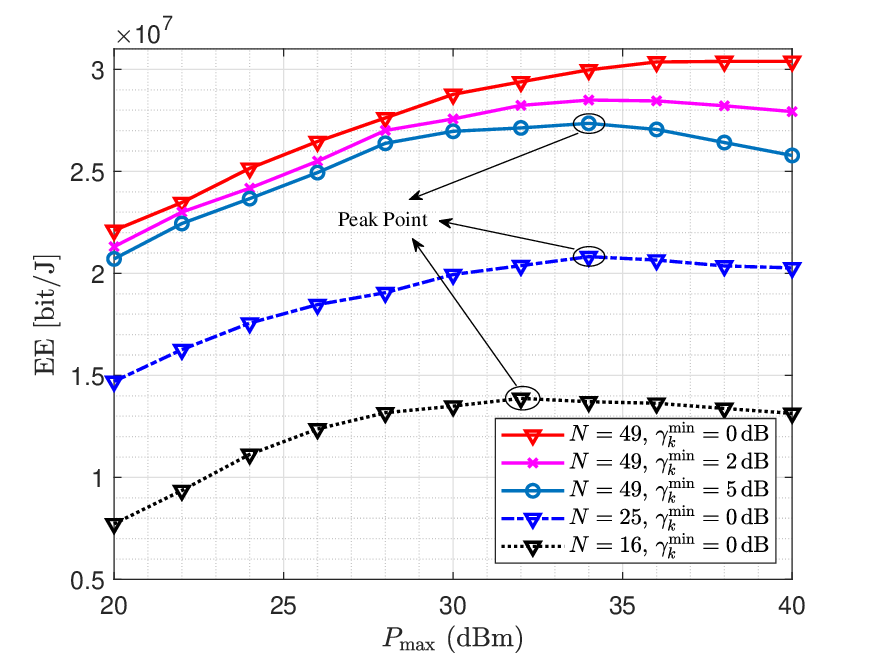}
\caption{Average EE against the number of SIM meta-atoms per layer for different values of ${{P}_{\rm{max}}}$ and ${{{\gamma}^{\rm{min}}_{k}}}$ considering the SDP algorithm and RF chain selection ($L = 4$, $M = 4$, $K = 4$, ${{{P}^{\rm{max}}_{l}}} = 30 \,\rm{dBm}$, ${{P}_{\rm{meta}}} = 10 \,\rm{dBm}$).}\label{Fig_N_Pmax_EE}\vspace{-0.2cm}
\end{figure}
Fig.~\ref{Fig_N_Pmax_EE} illustrates the average EE against the number of SIM meta-atoms per layer considering different values of ${{P}_{\rm{max}}}$ and ${{{\gamma}^{\rm{min}}_{k}}}$. As the maximum transmission power of the BS increases, the EE initially increases and then decreases. This behavior can be attributed to the impact of $P_{\text{max}}$ on the power consumption. Initially, enhancing the transmission power significantly boosts the sum SE, thus increasing the EE. However, at higher power levels, incremental further $P_{\text{max}}$ results only in minimal improvements in the sum SE while leading to a disproportionate increase of the energy consumption, thereby diminishing the overall EE.
Also, when the number of SIM meta-atoms is fixed, an increase in the required minimum rate per user leads to a decrease of the EE. Moreover, the larger ${\gamma^{\rm{min}}_{k}}$ (minimum QoS), the more significant the decline of the EE. This is because the BS needs to provide additional transmission energy to meet the rate requirements of the users, thereby increasing the total energy consumption.

\begin{figure}[t]
\centering
\includegraphics[scale=0.5]{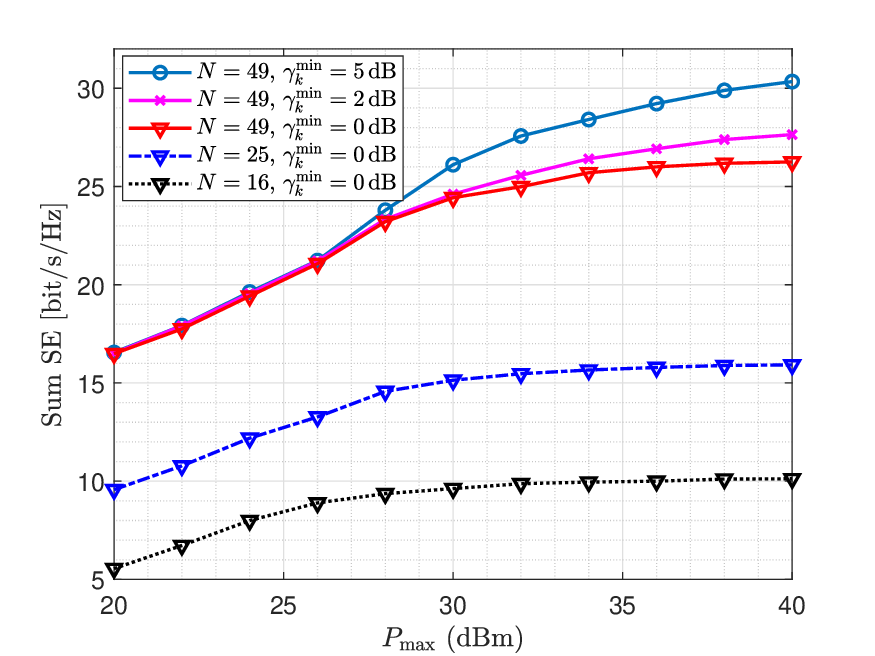}
\caption{Average sum SE against the number of SIM meta-atoms per layer for different values of ${{P}_{\rm{max}}}$ and ${{{\gamma}^{\rm{min}}_{k}}}$ considering the SDP algorithm and RF chain selection ($L = 4$, $M = 4$, $K = 4$, ${{{P}^{\rm{max}}_{l}}} = 30 \,\rm{dBm}$, ${{P}_{\rm{meta}}} = 10 \,\rm{dBm}$).}\label{Fig_N_Pmax_SE}\vspace{-0.2cm}
\end{figure}
Fig.~\ref{Fig_N_Pmax_SE} shows the average sum SE against the number of SIM meta-atoms per layer considering different values of ${{P}_{\rm{max}}}$ and ${{{\gamma}^{\rm{min}}_{k}}}$. The results show that when the minimum QoS is constant, i.e., ${\gamma^{\rm{min}}_{k}} = 0\,\rm{dB}$, increasing the maximum transmission power of the BS is always beneficial, but with diminishing returns when the transmission power is sufficiently large. Moreover, as the maximum transmission power increases, the impact of the minimum rate constraint becomes increasingly significant. Specifically, in scenarios where the maximum transmission power's ${{P}^{\rm{max}}}<28 \,\rm{dBm}$, the minimum rate requirement has virtually no impact on the system sum SE. Figs.~\ref{Fig_N_Pmax_EE} and~\ref{Fig_N_Pmax_SE} reveal that as $P_{\rm{max}}$ increases, the enforcement of stricter QoS constraints results in an earlier turning point for the decline of the EE due to the allocation of additional BS transmission power to meet the rate requirements of the UE.

\begin{figure}[t]
\centering
\includegraphics[scale=0.5]{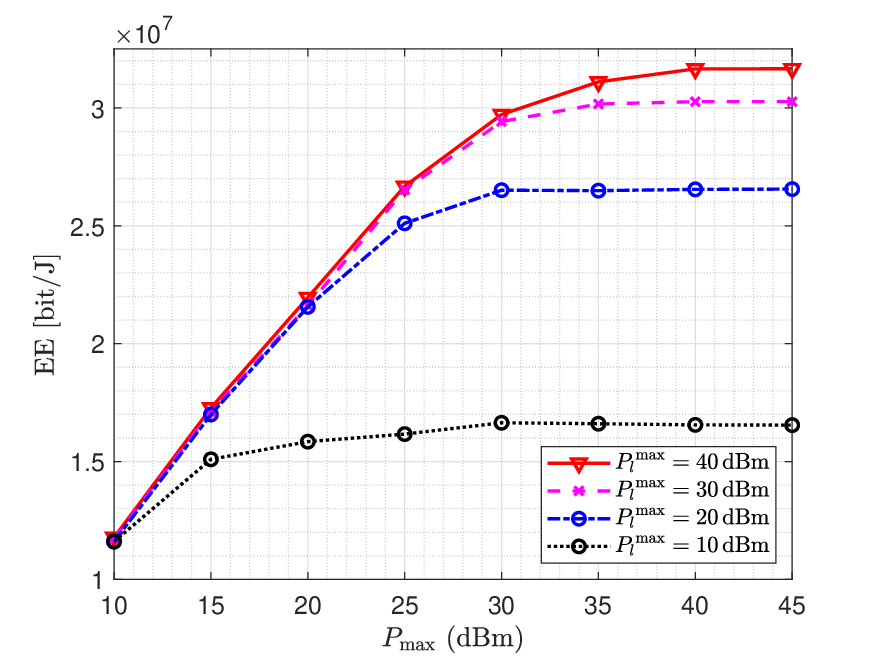}
\caption{Average EE against different values of ${{P}_{\rm{max}}}$ and ${{{P}^{\rm{max}}_{l}}}$ considering the SDP algorithm and RF chain selection ($L = 4$, $M = 4$, $K = 4$, $N = 49$, ${{P}_{\rm{meta}}} = 10 \,\rm{dBm}$, ${{{\gamma}^{\rm{min}}_{k}}} = 0 \,\rm{dB}$).}\label{Fig_Pmax_Pantenna_EE}\vspace{-0.2cm}
\end{figure}
Fig.~\ref{Fig_Pmax_Pantenna_EE} illustrates the relationship between the average EE and the maximum transmit power $P_{\text{max}}$ considering different values of $P_{l}^{\text{max}}$. The SDP algorithm with RF chain selection is considered.
The results reveal that as $P_{\text{max}}$ increases, the EE initially grows significantly across all $P_{l}^{\text{max}}$ configurations, but the rate of improvement decreases and eventually saturates.
However, the corresponding saturation point varies for different values for the maximum transmission power of the antenna. For example, when $P_{l}^{\text{max}} = 10 \, \text{dBm}$, the performance hardly improves after the BS maximum transmission power reaches 15 $\rm{dBm}$. In contrast, when $P_{l}^{\text{max}} = 20 \, \text{dBm}$, the performance stabilizes after the maximum transmission power reaches 30 $\rm{dBm}$. This is because when the BS transmission power is relatively low, the antenna transmission power does not limit the system performance. However, when the BS transmission power is high, each antenna cannot provide the corresponding transmission power, thus limiting the improvement in system performance.
Also, for $P_{l}^{\text{max}} = 40 \, \text{dBm}$, the highest EE is achieved, indicating that larger values of $P_{l}^{\text{max}}$ result in higher EE. The results indicate that the ratio between the antenna transmission power and the total BS transmission power is a critical parameter for ensuring good communication performance while minimizing hardware costs and energy consumption.


\begin{figure}[t]
\centering
\includegraphics[scale=0.5]{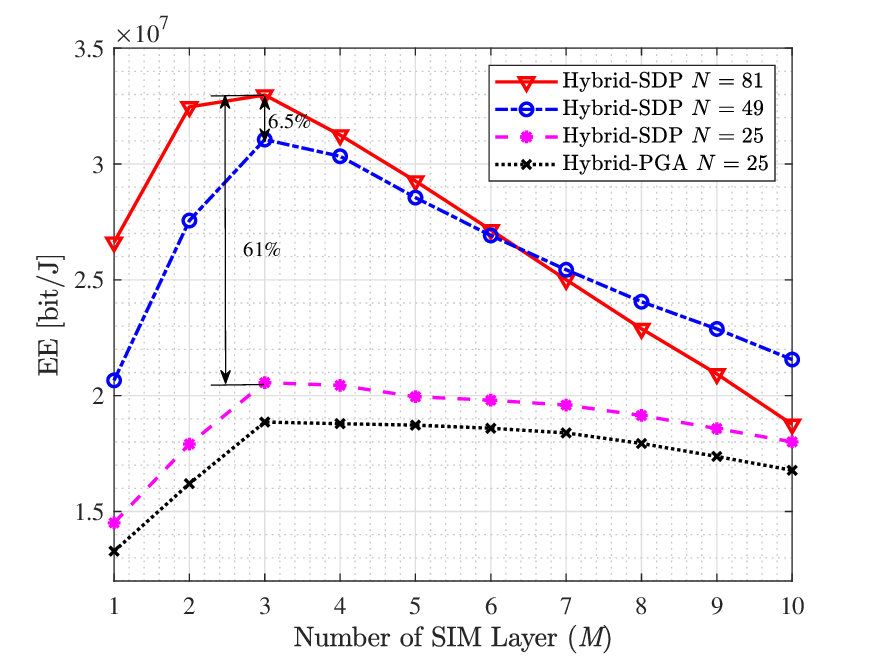}
\caption{Average EE against the number of SIM layers and the number of meta-atoms per layer considering the SDP algorithm and RF chain selection ($L = 4$, $K = 4$, ${{P}_{\rm{max}}} = 35 \,\rm{dBm}$, ${{{P}^{\rm{max}}_{l}}} = 30 \,\rm{dBm}$, ${{P}_{\rm{meta}}} = 10 \,\rm{dBm}$, ${{{\gamma}^{\rm{min}}_{k}}} = 0 \,\rm{dB}$).}\label{Fig_M_N_EE}\vspace{-0.2cm}
\end{figure}
Fig.~\ref{Fig_M_N_EE} illustrates the average EE against the number of SIM layers $M$ and the number of meta-atoms $N$ per layer. The SDP algorithm with RF chain selection is considered. We see that as $M$ increases, the EE first increases and then decreases. Furthermore, at the peak, when $N = 81$, the EE improves by 6.5\% and 61\% compared to $N = 49$ and $N = 25$, respectively. That is, for the EE, an optimal number of layers exists that provides the highest EE. In the considered scenario, this value is 3 layers. Moreover, when the number of SIM layer's $M>3$, the larger the number of meta-atoms per layer, the more significant the degradation in EE. When $M>6$ and $N = 81$, further increasing the number of layers results in a worse EE than that for $N = 49$. Therefore, if the EE is the primary metric, blindly increasing the number of layers is not a reasonable choice. It is essential to design the number of layers based on the number of meta-atoms per layer to achieve optimal performance.

\begin{figure}[t]
\centering
\includegraphics[scale=0.5]{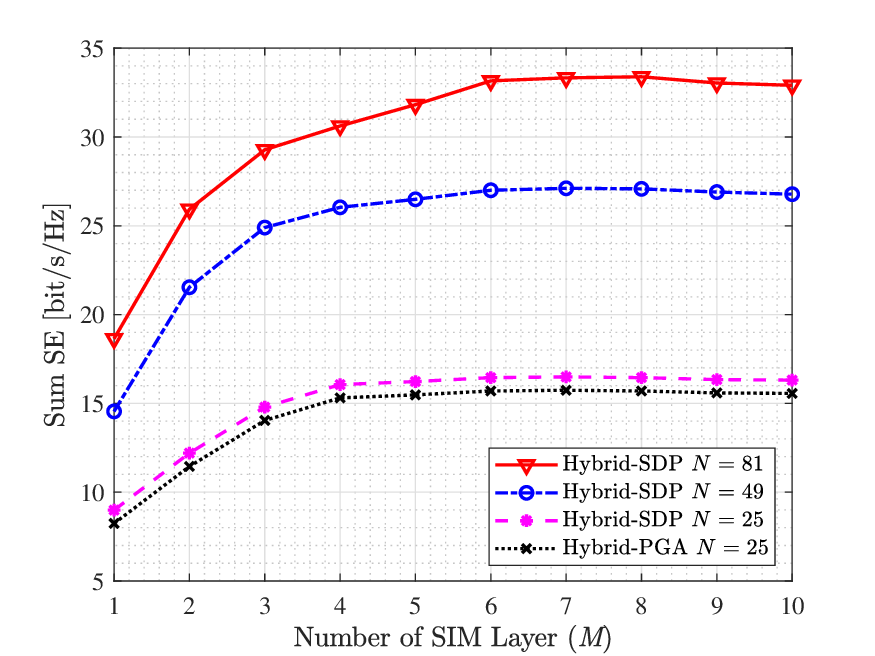}
\caption{Average sum SE against the number of SIM layers and the number of meta-atoms per layer considering the SDP algorithm and RF chain selection ($L = 4$, $K = 4$, ${{P}_{\rm{max}}} = 35 \,\rm{dBm}$, ${{{P}^{\rm{max}}_{l}}} = 30 \,\rm{dBm}$, ${{P}_{\rm{meta}}} = 10 \,\rm{dBm}$, ${{{\gamma}^{\rm{min}}_{k}}} = 0 \,\rm{dB}$).}\label{Fig_M_N_SE}\vspace{-0.2cm}
\end{figure}
Fig.~\ref{Fig_M_N_SE} shows the average sum SE against the number of SIM layers $M$ and the number of meta-atoms $N$ per layer. The SDP algorithm with RF chain selection is considered. The results show that when the number of SIM layers is small, increasing the number of SIM layers significantly improves system performance. However, when the number of layers exceeds a certain threshold, further increasing it does not continue to enhance the sum SE and may even lead to a decline. 
This is due to the product channel model in \eqref{G_l} and the distance-dependent inter-layer channels in \eqref{w}.
This outcome is consistent with the results shown in Fig.~\ref{Fig_M_N_EE}. Therefore, more SIM layers do not necessarily improve the performance, and the obtained results suggest that an optimal value to obtain satisfactory EE and SE is 2-5 layers, based on the considered system setup and parameters.

\begin{figure}[t]
\centering
\includegraphics[scale=0.5]{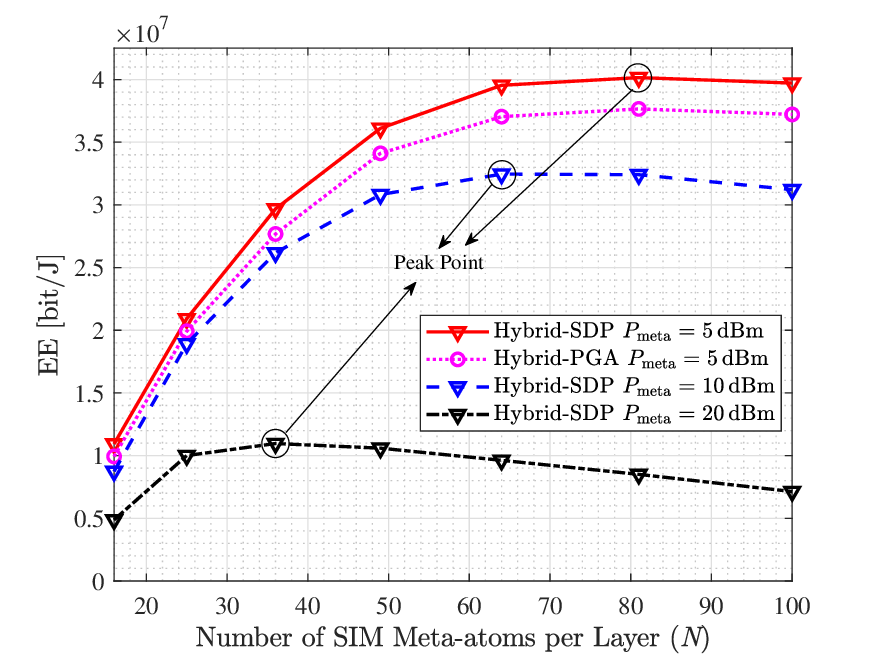}
\caption{Average EE against the number of SIM meta-atoms per layer for different values of ${P}_{\rm{meta}}$ considering the SDP/PGA algorithms and RF chain selection ($L = 4$, $M = 4$, $K = 4$, ${{P}_{\rm{max}}} = 35 \,\rm{dBm}$, ${{{P}^{\rm{max}}_{l}}} = 30 \,\rm{dBm}$, ${{{\gamma}^{\rm{min}}_{k}}} = 0 \,\rm{dB}$).}\label{Fig_N_Pmeta_EE}\vspace{-0.2cm}
\end{figure}
Fig.~\ref{Fig_N_Pmeta_EE} illustrates the average EE against the number of SIM meta-atoms per layer $N$ for different values of power consumption of each meta-atom by considering the SDP/PGA algrithms and RF chain selection. The results show that when the number of meta-atoms per layer is small, increasing the number of meta-atoms leads to the improvement of EE performance. However, after reaching a peak, further increasing the number of meta-atoms leads to the degradation of EE. Moreover, it is evident that the larger the power consumption per meta-atom, the worse the EE. Additionally, as the number of meta-atoms increases, the gap of EE with different meta-atom powers becomes more pronounced. For example, the EE with each meta-atom at 5 $\rm{dBm}$ improves by 170\% compared to 20 $\rm{dBm}$ when $N = 36$ and by 400\% when $N = 81$. Thus, minimizing the power consumption of SIM meta-atoms is always beneficial for improving the system EE performance.


\begin{figure*}[t!]
\normalsize
\setcounter{mytempeqncnt}{1}
\setcounter{equation}{53}
\begin{align}\label{Them4}
\frac{{\partial {{\left| {{\bf{h}}_{{\rm{SIM}},k}^{\rm{H}}{\bf{G}}{{\bf{W}}_1}{{\bf{p}}_j}} \right|}^2}}}{{\partial \varphi _m^n}} &= \frac{{\partial {{\left| {\sum\limits_{n = 1}^N {{e^{j\varphi _m^n}}} {\bf{h}}_{{\rm{SIM}},k}^{\rm{H}}{\bf{b}}_m^n{{\left( {{\bf{q}}_m^n} \right)}^{\rm{H}}}{{\bf{W}}_1}{{\bf{p}}_j}} \right|}^2}}}{{\partial \varphi _m^n}}= \frac{{\partial \Re \left[ {\left( {{e^{j\varphi _m^n}}{\bf{h}}_{{\rm{SIM}},k}^{\rm{H}}{\bf{b}}_m^n{{\left( {{\bf{q}}_m^n} \right)}^{\rm{H}}}{{\bf{W}}_1}{{\bf{p}}_j}} \right){{\left( {{\bf{h}}_{{\rm{SIM}},k}^{\rm{H}}{\bf{G}}{{\bf{W}}_1}{{\bf{p}}_j}} \right)}^{\rm{H}}}} \right]}}{{\partial \varphi _m^n}}\notag\\
 &= \Im \left[ {{{\left( {{e^{j\varphi _m^n}}{\bf{h}}_{{\rm{SIM}},k}^{\rm{H}}{\bf{b}}_m^n{{\left( {{\bf{q}}_m^n} \right)}^{\rm{H}}}{{\bf{W}}_1}{{\bf{p}}_j}} \right)}^{\rm{H}}}\left( {{\bf{h}}_{{\rm{SIM}},k}^{\rm{H}}{\bf{G}}{{\bf{W}}_1}{{\bf{p}}_j}} \right)} \right]
\end{align}
\setcounter{equation}{54}
\hrulefill
\end{figure*}

\section{Conclusion}
In this paper, we investigated the integration of SIM into a multi-antenna BS for downlink multi-user communications, focusing on EE optimization. We formulated a hybrid precoding optimization problem considering BS antenna precoding and SIM wave-based beamforming design. Given the non-convex nature of the problem, we proposed an AO-based framework to obtain a suboptimal solution. Specifically, we employed the SCA algorithm for BS precoding and utilized two distinct algorithms, SDP and PGA, for SIM beamforming.
The results revealed that the proposed hybrid precoding achieves up to an 80\% improvement in EE compared to digital precoding without SIM beamforming.
Finally, the results demonstrate that, despite the discrepancy in the optimal number of SIM layers for maximizing the EE and SE, selecting between 2 and 5 layers offers a good compromise, yielding satisfactory performance in terms of both metrics.

\begin{appendices}
\section{Proof of Theorem 1}
This appendix reports the proof of \eqref{deta_R} of Theorem \ref{them}. To start with, we note that the gradient of $R_{\rm{sum}}$ with respect to $\varphi^n_{m}$ can be written as
\begin{align}\label{Them1}
\setcounter{equation}{49}
    \frac{{\partial {R_{{\rm{sum}}}}}}{{\partial \varphi _{m}^n}} \!=\! {\log _2}e \!\cdot\!\! \sum\limits_{k = 1}^K {\frac{1}{{1 + {\gamma _k}}} \cdot } \frac{{\partial {\gamma _k}}}{{\partial \varphi _{m}^n}},\forall m \in {\cal M},\forall n \in {\cal N}.
\end{align}
Based on the standard quotient rule for derivatives, we compute $\frac{{\partial {\gamma _k}}}{{\partial \varphi _{m}^n}}$ in \eqref{Them1} as
\begin{align}\label{Them2}
&\frac{{\partial {\gamma _k}}}{{\partial \varphi _m^n}} = \frac{1}{{{\zeta _k}}} \cdot \frac{{\partial {{\left| {{\bf{h}}_{{\rm{SIM}},k}^{\rm{H}}{\bf{G}}{{\bf{W}}_1}{{\bf{p}}_k}} \right|}^2}}}{{\partial \varphi _m^n}}\notag\\
&- \frac{{{{\left| {{\bf{h}}_{{\rm{SIM}},k}^{\rm{H}}{\bf{G}}{{\bf{W}}_1}{{\bf{p}}_k}} \right|}^2}}}{{\zeta _k^2}} \cdot \mathop \sum \limits_{j = 1,j \ne k}^K \frac{{\partial {{\left| {{\bf{h}}_{{\rm{SIM}},k}^{\rm{H}}{\bf{G}}{{\bf{W}}_1}{{\bf{p}}_k}} \right|}^2}}}{{\partial \varphi _m^n}},
\end{align}
where ${\xi _k}$ is given by
\begin{align}\label{xi_k}
{\zeta _k} = \mathop \sum \limits_{j = 1,j \ne k}^K {\left| {{\bf{h}}_{{\rm{SIM}},k}^{\rm{H}}{\bf{G}}{{\bf{W}}_1}{{\bf{p}}_j}} \right|^2} + \sigma _k^2.
\end{align}

With the help of \eqref{Them2}, the partial derivative $\frac{{\partial {R_{{\rm{sum}}}}}}{{\partial \varphi _{m}^n}}$ in \eqref{Them1} can be computed and simplified as
\begin{align}\label{Them3}
\frac{{\partial {R_{{\rm{sum}}}}}}{{\partial \varphi _m^n}} = {\log _2}e \cdot \sum\limits_{k = 1}^K {{\chi _k}}  \cdot \left( {\frac{{\partial {{\left| {{\bf{h}}_{{\rm{SIM}},k}^{\rm{H}}{\bf{G}}{{\bf{W}}_1}{{\bf{p}}_k}} \right|}^2}}}{{\partial \varphi _m^n}}} \right.\notag\\
\left. { - {\gamma _k}\mathop \sum \limits_{j = 1,j \ne k}^K \frac{{\partial {{\left| {{\bf{h}}_{{\rm{SIM}},k}^{\rm{H}}{\bf{G}}{{\bf{W}}_1}{{\bf{p}}_j}} \right|}^2}}}{{\partial \varphi _m^n}}} \right),
\end{align}
where ${\chi _k} = \frac{1}{{{\zeta _k}}}$ is defined in \eqref{dalta}.

The primary challenge of computing \eqref{Them3} is determining the partial derivative of ${{{\left| {{\bf{h}}_{{\rm{SIM}},k}^{\rm{H}}{\bf{G}}{{\bf{W}}_1}{{\bf{p}}_k}} \right|}^2}}$ with respect to ${e^{j\varphi _{m}^n}}$. It is important to note that ${{\bf{h}}_{{\rm{SIM}},k}^{\rm{H}}{\bf{G}}{{\bf{W}}_1}{{\bf{p}}_k}}$ is linear with respect to ${e^{j\varphi _{m}^n}}$. Therefore, for each pair $k,j \in \mathcal{K}$, we obtain \eqref{Them4} shown at the top of this page, where ${{\bf{b}}_{m}^n}$ and ${{{( {{\bf{q}}_{m}^n} )}^{\rm{H}}}}$ are defined in \eqref{B} and \eqref{Q}, respectively.
Finally, the proof is completed by substituting \eqref{Them4} into \eqref{Them3}.
\end{appendices}

\bibliographystyle{IEEEtran}
\bibliography{IEEEabrv,Ref}

\end{document}